\documentclass[fleqn,usenatbib]{mnras}

\usepackage{newtxtext,newtxmath}

\usepackage[T1]{fontenc}
\usepackage{ae,aecompl}
\usepackage[utf8]{inputenc}
\usepackage{graphicx}	
\usepackage{rotating}
\usepackage{amsmath}	
\usepackage{amssymb}	

\newcommand{\etal}{et al.\ }

\newcommand{\ha}{H$\alpha$}
\newcommand{\re}{$\text{R}_\text{e}$}
\newcommand{\HII}{\ion{H}{ii}}

\newcommand{\piped}{{\sc Pipe3D}}

\title[Outlying \ha{} emitters in SDSS IV MaNGA]{Outlying \ha{} emitters in SDSS IV MaNGA}

\author[Bait \etal]
{Omkar Bait,$^{1}$\thanks{E-mail: omkar@ncra.tifr.res.in (OB)}, 
Yogesh Wadadekar$^{1}$\thanks{E-mail: yogesh@ncra.tifr.res.in
  (YW)}, Sudhanshu Barway,$^{2}$\thanks{E-mail: sudhanshu.barway@iiap.res.in (SB) }
    \\
$^{1}$National Centre for Radio Astrophysics, Tata Institute of Fundamental Research, Post Bag 3, Ganeshkhind, Pune
411007, India \\
$^{2}$Indian Institute of Astrophysics (IIA), II Block, Koramangala, Bengaluru 560 034, India}

\date{Accepted XXX. Received YYY; in original form ZZZ}

\pubyear{2018}

\begin{document}

\pagerange{\pageref{firstpage}--\pageref{lastpage}}
\maketitle

\begin{abstract}

We have carried out a systematic search for outlying \ha{} emitters in the entire data release 14 of the Sloan Digital Sky Survey (SDSS) IV Mapping Nearby Galaxies at APO (MaNGA) survey. We have discovered six outlying \ha{} emitters with no bright underlying optical continuum emission in the imaging data release 5 from the Dark Energy Camera Legacy Survey (DECaLS) and data release 6 of the Mayall $z$-band Legacy Survey (MzLS) + Beijing-Arizona Sky Survey (BASS). They also show a velocity field which is different from that of the host galaxy. These outlying \ha{} emitters all have extended structure in the \ha{} image. Their emission line ratios show that they are photoionised due to an active galactic nucleus (AGN) or a mixture of both an AGN and star formation. Some of them are  very likely to be fainter counterparts of Hanny's Voorwerp like objects.

\end{abstract}


\section{Introduction}

Large-scale galactic outflows due to both starburts and active galactic nuclei (AGN) are quite common, particularly at high redshifts, and play an important role in providing feedback in galaxy formation and evolution \citep[see][for a review]{Veilleux05}. Such galaxy outflows can appear as \ha{} emission away from the galaxy. The classic example is that of Messier 82 (M82) which shows large scale bipolar superwinds \citep{Bland88, Shopbell98} due to the ongoing starburst at the centre. Interestingly, M82 also shows an \ha{} cap like structure at a projected distance of about 11 kpc from the disc \citep{Devine99, Lehnert99}. Typically such galactic outflows are known to trigger star formation in the host galaxy as a consequence of gas compression. However, \cite{Maiolino17} has  also found observational evidence for star formation occurring in the galactic outflow at a rate of more than 15 solar masses per year.

Another interesting case of  an outlying emission line region is that of the Hanny's Voorwerp \citep{Lintott09}, discovered serendipitously,  lying near the spiral galaxy IC 2497. This object was discovered as part of the Galaxy Zoo project in the Sloan Digital Sky Survey  (SDSS) $g$-band image due to its strong emission in the [OIII]4959, 5007 lines. It has been suggested that the gas in Hanny's Voorwerp has a tidal origin \citep{Josza09}. This gas was then photoionised by the nucleus of IC 2497, which, in turn, was a quasi-stellar object (QSO) that has faded dramatically in the last $10^5$ years, thus producing a quasar light echo \citep{Lintott09, Schawinski10}. In \citet{Keel12b}, Galaxy Zoo volunteers have carried out further searches to identify 19 more such objects, which are termed as extended emission-line regions (EELRs). Of these, most are in interacting or merging state which further makes the tidal origin of gas which was photoionised by an AGN a more plausible explanation of EELRs.

With the advent of larges scale integral field unit (IFU) surveys e.g., the Mapping Nearby Galaxies at Apache Point Observatory \citep[][MaNGA]{Bundy2015}, Calar Alto Legacy Integral Field Area survey \citep[][CALIFA]{Sanchez12}, and the Sydney-AAO Multi-object Integral-field spectrograph Galaxy survey \citep[][SAMI]{Bryant15} we can now search for such outlying \ha{} emitters around several thousand  galaxies. In the MaNGA survey, \citet{Cheung16b}  found an outlying \ha{} complex without any optical counterpart about 6.3 kpc away from the  effective radius of the host galaxy and at about 3 times the effective radius of the galaxy. Using a detailed analysis of the gas kinematics, gas phase metallicity, and emission-line ratios the authors argue that the scenario is most consistent with an gas accretion event. Recently, \citet{Lin17} have discovered a large outlying \ha{} blob ($\sim$ 3-4 kpc in radius) around 8 kpc away from the host galaxy which is undergoing a dry galaxy merger. Here the authors argue that this could either be an AGN outburst or a possible ``ultra-diffuse galaxy". Using the SAMI survey, \citet{Fogarty12} have discovered an outlying emission-line region off the plane of the disk of edge-on galaxy ESO 185-G031 which is ionised due to shock excitation from a starburst driven wind. \citet{Richards14} have found a large \HII{} complex in the dwarf galaxy GAMA J141103.98--003242.3 which is similar to the Large Megallanic Cloud (LMC) and 30 Doradus complex. Recently, using the Multi Unit Spectroscopic Explorer (MUSE), \citet{Epinat18} has found a large diffuse ionised gas complex ($\sim 10^4\ $kpc$^2$ in area) in [OII]$\lambda\lambda$3727, 3729 in a galaxy group COSMOS-Gr30 at a redshift of 0.725 with two kinematic sub-structures. The authors argue that atleast one of the gas component has a non-primordial origin arising either due to tidal interactions between the group galaxies or gas expelled due to AGN outflows. More recently, \citet{LopezCoba18} have conducted a systematic search for galaxy outflows in the Local Universe using the CALIFA survey and have found 17 objects which show evidence of galactic winds. They find that these objects do not show any excess in the their integrated star formation rates (SFRs) compared to  normal star forming galaxies. However, the SFR density in the inner regions shows an excess and is an important driver of galaxy outflows.

Thus a systematic study of outlying \ha{} emitters is important from the point of view of understanding galaxy outflows and interestingly also in-situ star formation in these extreme environments. With a view to increase the sample size of these interesting but rare class of objects, we have carried a systematic search of outlying \ha{} emitters in the recently released SDSS IV MaNGA IFU survey. We use the recently released MaNGA value added catalogue, analysed using the Pipe3D pipeline \citep{Sanchez16b, Sanchez18}, of 2,755 galaxies (see Sec. \ref{sample} for details on the sample selection). We visually identify outlying \ha{} emitting regions around MaNGA target galaxies on which the IFU is centered. These emission line regions also shows a different velocity component from that of the host galaxy. Note that our search criteria differ from those of \citet{LopezCoba18}, where they have used various diagnostic diagrams (e.g., EW(\ha{}), [NII]/\ha{} line ratios etc.) which are tuned to find galactic winds, whereas we are interested in finding extended emission line regions away from the MaNGA target galaxies. We have found a total of six such outlying 
\ha{} emitters which have extended \ha{} structure around the host galaxies (see Sec. \ref{sample}). Interestingly using the Baldwin Philip and Terlevich (BPT) diagram, we find that the extended \ha{} emitters shows various sources of ionisation (see Section \ref{BPT}). In section \ref{comparison} we briefly discuss the possibility that these extended emitters are low luminosity counterparts of Hanny's Voorwerp \citep{Lintott09} like objects. In section \ref{notes} we briefly discuss some of the interesting sources in our sample. We will then summarise and conclude in Sec \ref{conclusion}.  
       
Throughout this paper we use the standard concordance
cosmology with $\Omega_M = 0.3$, $\Omega_\Lambda = 0.7$ and $h_{100} =
0.7$.

\section{Sample Selection} \label{sample}
Our sample is drawn from Mapping Nearby Galaxies at Apache Point Observatory \citep[][MaNGA]{Bundy2015} survey which was released as part of the  Sloan Digital Sky Survey Data Release 14 \citep[][SDSS DR14]{SDSS14}. This recent data release contains a sample of 2,812 MaNGA datacubes. These datacubes were analysed using the Pipe3D data analysis pipeline by the MaNGA team and their final catalogue of 2,755 galaxies forms our sample, wherein we search for outlying \ha{} emitters.

\subsection{The MaNGA survey}
MaNGA is an integral field unit (IFU) survey which aims to observe a total sample of $\sim 10,000$  nearby galaxies. The MaNGA instrument  consists of 17 hexagonal shaped science IFUs mounted on the 2.5 metre Apache Point Observatory which has a 3$^\circ$ field of view. The IFU sizes range from 19 fibres with 12\arcsec on-sky diameter to 127 fibres with 32\arcsec on-sky diameter. Additionally, there are 12 mini-bundles with 7 fibres each which are used for spectrophotometric calibration, and 92 single fibres used for sky subtraction. The details of the MaNGA instrument are described in \citet{Drory2015}. The MaNGA fibres are then fed to the two Baryon Oscillation Sky Survey (BOSS) spectrographs \citep{Smee2013}, each of which contain a red and a blue camera which together have a wavelength coverage from 3600 \AA \ to 10300 \AA \ and a spectral resolution, R $\sim  2000$. The integration time for each of the MaNGA targets depends on the signal to noise ratio requirement of 5 \AA$^{-1}$ fibre$^{-1}$ in $r$-band continuum at a surface brightness of 23 AB arcsec$^{-2}$ \citep{Law2015}. The raw data are then reduced using the MaNGA Data Reduction Pipeline (DRP) as described in \citet{Law2016}, which constructs the sky subtracted and flux calibrated datacubes for each of the MaNGA galaxies. The MaNGA PSF can be well described by a single Gaussian with median spatial resolution of $\sim$2.54 arcsec FWHM. The median instrumental velocity resolution is 72 kms$^{-1}$. The typical relative flux calibration errors are 1.7\% between \ha{} and H$\beta$, and about 4.7\% between [N \i\i] $\lambda$6583 and [O \i\i] $\lambda$3727 \citep{Yan2016}. 

The primary MaNGA selection criteria produce a sample of roughly 10,000 galaxies which is complete above the stellar mass limit of $ 10^9\ \text{M}_\odot$ and has a roughly flat stellar mass distribution \citep[See][]{Bundy2015, Wake2017}. More constraints are put in order to maximise spatial coverage and S/N which favours a lower redshift sample. Other constraints favour a higher redshift sample. Taking these constraints into account the MaNGA target galaxies are divided into two classes: Primary and Secondary samples. The Primary sample has a mean redshift of 0.03 and reaches 1.5 times the galaxy half-light radius (\re{}) for roughly 80\% of its targets. The Secondary sample has a slightly higher mean redshift of 0.045 and reaches 2.5 \re{} for roughly 80\% of its targets. Of the total target sample of $\sim$10,000 galaxies, the data release 14 contains a sample of 2,812 reduced MaNGA datacubes. 

\subsection{\piped{} value added catalogue}
In this study, we use the MaNGA value added catalogue analysed using the \piped{} pipeline \citep{Sanchez16b, Sanchez18} which is based on {\sc FIT3D}, the details of which are described in \citet{Sanchez16a}. We briefly describe the fitting algorithm here. 

{\sc FIT3D} follows a two level modelling of the integral field spectroscopy (IFS) data. At the first level, it models the stellar kinematics, stellar population and dust attenuation using the continuum. At this level, initially the stellar kinematics (only the velocity and dispersion) are estimated assuming a Gaussian profile\footnote{{\sc FIT3D} does not model higher order kinematics as it is primarily developed for lower spectral resolution (R < 2000) IFS.}. This is achieved by first fixing the velocity dispersion and varying the velocity from the minimum to the maximum value until a best fit velocity is found. Then the velocity is fixed to the best fit value and the velocity dispersion is varied till the best-fit value is found. Following this, the dust attenuation is varied until the best fit dust attenuation is found, while keeping the velocity and velocity dispersion fixed to their best fit values.  Once the best fit value of these three parameters are found {\sc FIT3D} performs the stellar population synthesis (SPS) modelling using the GSD16 stellar template library \citep{Cid_Fernandes13}. GSD16 contains a total of 156 templates with 39 stellar ages, ranging from 1 Myr to 13 Gyr, and 4 metallicities: 0.2, 0.4, 1, and 1.5 times solar metallicity. Each of these templates are shifted according to the velocity, convolved with the velocity dispersion and dust attenuated, following the dust attenuation law of \citet{Cardelli89}, using the best fit values of the individual parameters. The average properties of the stellar population (e.g., stellar mass, light/mass weighted stellar age, metallicity, mass-to-light ratio, etc.) is derived using the weights on the individual single stellar population.

The second level of {\sc FIT3D} deals with the measurement of nebular emission lines. The emission line only spectrum is derived by subtracting the modeled stellar continuum obtained from the previous level. The uncertainities involved in the stellar population modelling are propagated to the emission line-only spectrum. Thereafter, around each emission line a single Gaussian and a low order polynomial is fitted. Following a similar procedure as in the previous level, the velocity, velocity dispersion and the intensity of each line is derived starting by first fitting for the velocity and fixing the other two parameters. This is followed by fitting for the velocity dispersion and finally for the intensity of the emission line. Furthermore, the \ha{} emission line is corrected for dust attenuation using the Balmer decrement for every spaxel assuming a \ha{}/H$\beta$ line ratio of 2.86 and adopting a \citet{Cardelli89} dust attenuation law and specific dust attenuation of R$_V$ of 3.1.

The current data release of the \piped{} value added catalogue\footnote{\url{http://www.sdss.org/dr14/manga/manga-data/manga-pipe3d-value-added-catalog/}} has provided dataproducts and catalogue for 2,755 galaxies in the MaNGA sample. 

\subsection{Optical Photometric data}

We also use the publicly available optical $g$, $r$, and $z$ band imaging data from the SDSS DR14, which have a 5$\sigma$ depths of 23.13, 22.7 and 20.71 respectively. We also use the data release 5 of the Dark Energy Camera Legacy Survey (DECaLS) and data release 6 of the Mayall z-band Legacy Survey (MzLS) + Beijing-Arizona Sky Survey (BASS). The DECaLS survey is being carried out on the Dark Energy Camera (DECam) on the Blanco 4m telescope at the Cerro Tololo Inter-American Observatory and it will provide the optical imaging for the Dark Energy Spectroscopic Instrument (DESI) footprint. DECam with its large field of view and good sensitivity can reach a 5$\sigma$ depths of 24.0, 23.4 and 22.5 in the $g$, $r$, and $z$ band respectively. The BASS survey is being carried out on the 2.3m Bok telescope using the 90Prime camera in the $g$ and $r$ band, and it will reach similar 5$\sigma$ depths of 24.0 and 23.4 in $g$ and $r$ band respectively. Similarly the MzLS survey is being carried out on the Mayall telescope at Kitt Peak National Observatory using the MOSAIC-3 camera only in the $z$-band and will reach a 5$\sigma$ depth of 23.0.

\subsection{Selection Criteria} 
\begin{figure*}
\begin{center}
\includegraphics[scale=0.5]{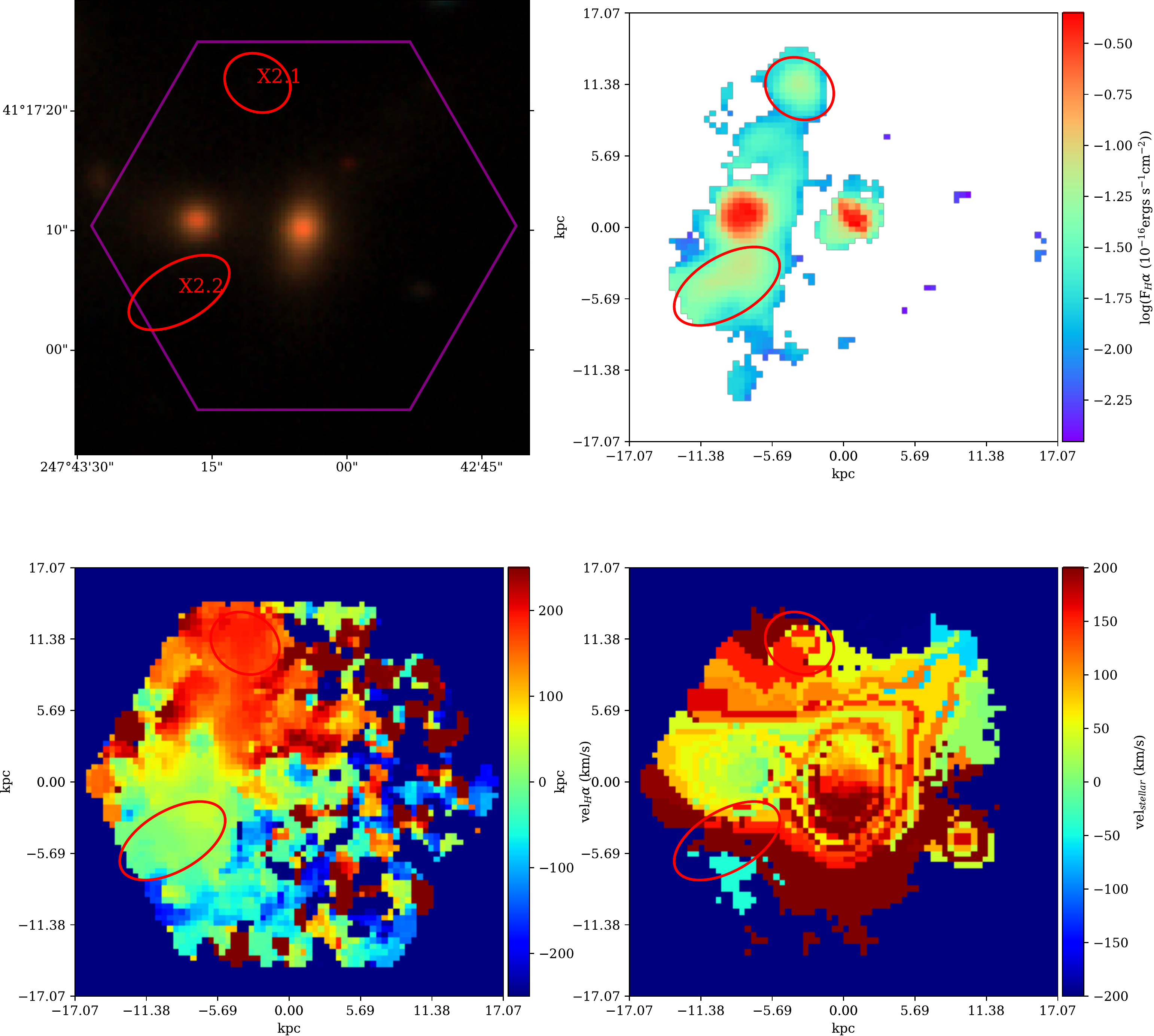}
\caption{MaNGA observation of a galaxy with MaNGA ID: manga-8601-12701. Top-left: MzLS/BASS grz \citet{Lupton04} colour composite image with the MaNGA IFU extent around the galaxy shown in purple in world coordinate system (WCS). Top-right: \ha{} flux map in logarithmic units. Bottom-left: \ha{} velocity map. Bottom-right: Stellar velocity map of the host galaxy. The outlying \ha{} emitters are circled in red in all the panels.  Notice that the \ha{} emitter does not have any bright optical counterpart, it also has a velocity component different from the host galaxy.} \label{example}
\end{center}
\end{figure*}

We identify a region of \ha{} emission as  an outlying \ha{} emitter if the following three criteria are satisfied:
\begin{enumerate}
\item A signal-to-noise ratio (SNR) of \ha{} flux greater than 5 and which covers atleast one MaNGA PSF which corresponds to $\sim$5 spaxels.
\item No bright optical counterpart in SDSS $gri$ colour composite image or in the DECaLS, MzLS/BASS $gri$ colour composite image. 
\item \ha{} velocity of the emitter which is different (typically upto 400 km/s) from the velocity field of the parent galaxy and/or  the \ha{} emission is clearly extra-planar.
\end{enumerate}

Under these criteria, we follow a fairly straightforward, two step approach, to visually search for such outlying \ha{} emitters. In the first step we select galaxies which satisfy the first two criteria. To do this, we produce panels for each galaxy in the MaNGA sample having the DECaLS/MzLS-BASS colour image and the \ha{} flux map. We visually inspect each galaxy using this panel and search for \ha{} emitting regions which do not show any bright optical counterpart. Starting from a sample of 2,755 MaNGA galaxies released in the \piped{} catalogue, we visually identify 133 galaxies which satisfy these two criteria. 

In the second step, we inspect the \ha{} velocity map of these shortlisted galaxies to make sure that the outlying \ha{} emission is different from that of the host galaxy. Fig. \ref{example}, shows an example of an outlying \ha{} emitter in our sample. Notice the large scale \ha{} emission away from the optical extent (first and second panel) of the host galaxy highlighted in red ellipses. The \ha{} velocity of the outlying emission is also different from that of the optical galaxy shown as seen from the third and fourth panels. Such a galaxy satisfies all the three selection criteria and is added in the sample of outlying \ha{} emitting region. We visually inspect the velocity map of all the 133 shortlisted galaxies in the first step of the selection and find a final sample of 41 galaxies which satisfy all the three selection criteria. 

\begin{figure*}
\begin{center}
\includegraphics[scale=0.6]{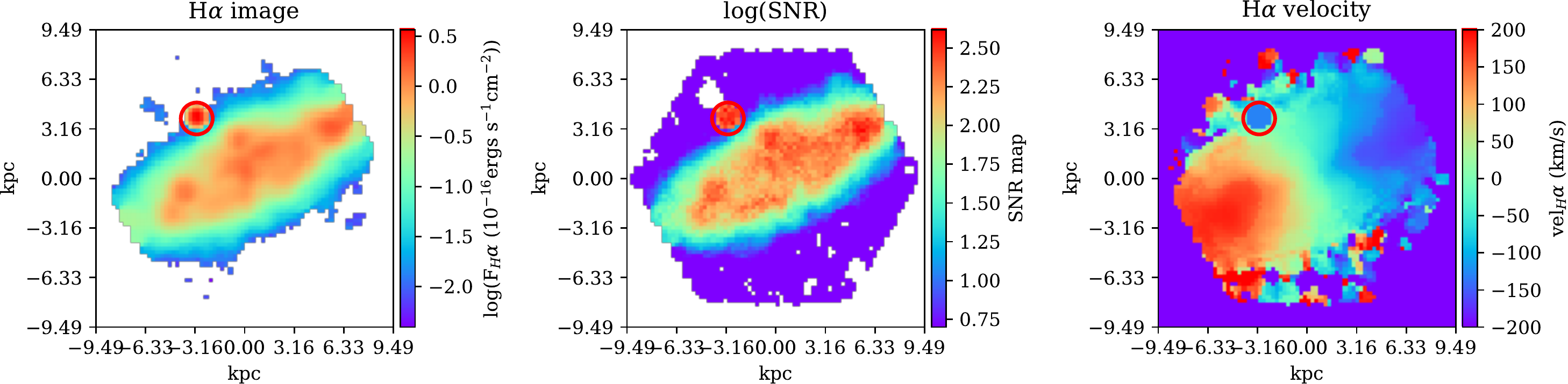}
\caption{MaNGA observation of a spiral galaxy with MaNGA ID: manga-8257-12703. Left panel: \ha{} flux map in logarithmic units. Middle panel: \ha{} logarithm of the signal to noise ratio with truncation below SNR of 5. Note that the apparent outlying \ha{} emitter which satisfies all the criteria and circled red is an artefact since it has a very narrow emission line width as seen in the MaNGA datacube (not shown here). The full list of such artefacts is given in the Appendix \ref{artefacts}. } \label{artefact}
\end{center}
\end{figure*}

We then individually inspect the spectrum of these outlying \ha{} emitters and find that 35 objects are artefacts since the FWHM of the \ha{} line is very narrow ($\sim $2 \AA) compared to the instrumental width of the MaNGA IFU. Of them about 29 of them appear like hot-pixels in the \ha{} image as shown in Fig. \ref{artefact}, and 6 have extended emission but a very narrow \ha{} FWHM. Notice that Fig. \ref{artefact}, shows the outlying emission with a very high SNR (as shown in red circle) and also has a very different velocity component from the host galaxy, however after visually inspecting the spectrum it is flagged as an artefact. The full list of artefacts in the MaNGA survey is given in the Appendix \ref{artefacts}, which is available in the online version of this paper.  Our final sample of outlying \ha{} emitters is summarised in Table \ref{final_sample}. We have also independently rediscovered the outlying \ha{} emitter from \citet{Lin17}, but do not discuss it here as it has been extensively studied in their paper.

\setcounter{table}{0}
\begin{table*}
\begin{center}
\caption{Final Sample of outlying \ha{} emitters in the SDSS MaNGA.}
\begin{tabular}{cccccccccc}
\hline \\
Obj ID &  RA & DEC & plate-ifu  & Host SDSS ObJ ID &RA (Host) &  DEC (Host) & z\\
        &     J2000 (hhmmss)    &  J2000 (ddmmss) &  & & J2000 (deg)  &                      J2000 (deg) & &  \\
\hline \\
X1  & 14:09:06.710 &+53:27:55.644 & 8591-6101 & 1237661416601878596 (HX1) &14:09:06.87 & +53:27:48.70 & 0.0785 \\
X2.1 & 16:30:52.685 & +41:17:22.074  & 8601-12701 & 1237655374109802661 (HX2) &16:30:53.11 & +41:17:10.81 & 0.0936 \\
X2.2 & 16:30:53.268 & +41:17:04.565  & 8601-12701 & 1237655374109802661 (HX2) &16:30:53.11 & +41:17:10.81  & 0.0936 \\
X3  & 08:47:47.430 & +54:01:29.807  & 8724-6102 & 1237651533336281291 (HX3)  & 08:47:46.76 & +54:01:36.02  & 0.0472 \\
X4  & 08:18:50.238 & +22:57:14.864  & 8939-12701 & 1237661085345513774 (HX4) & 08:18:49.61 & +22:57:16.37  & 0.0924 \\
X5  & 11:15:47.004 & +50:24:09.997  & 8947-3701 & 1237657855533056062 (HX5) & 11:15:47.46 & +50:24:05.75  & 0.0475 \\
\hline
\hline \\

\end{tabular} \label{final_sample}
\end{center}
Notes.\ Column (1) gives the object ID of the outlying \ha{} emitter,
column (2) and (3) gives the RA and DEC of the outlying \ha{} emitter respectively. Column (4) and (5) gives the MaNGA  plateifu name and SDSS DR14 object ID of the host galaxy respectively. And columns (6), (7) and (8) give the RA and DEC and SDSS spectroscopic redshift of the host galaxy respectively.  
\end{table*}

\section{Results and Discussions}
In this section, we measure the integrated emission line fluxes using aperture photometry of the outlying \ha{} emitters. We then study their position on the emission line ratio diagnostic diagram of \citet{BPT}. Finally, we individually describe the objects in our sample.

\subsection{Integrated properties of outlying \ha{} emitters}
In this section, we study the integrated emission line fluxes of the outlying \ha{} emitters from our sample. In particular, in order to know the source of ionisation we want to identify the location of these outlying \ha{} emitters on the ``BPT diagram" \citep{BPT}. Hence, we measure the total emission line flux in the \ha{}, [NII]$\lambda\ 6583$, [OIII]$\lambda\ 5007$, and the H$\beta$ lines. 

We make elliptical apertures \footnote{The aperture photometry was performed using the photutils package \url{https://photutils.readthedocs.io/en/stable/}.} around all the extended \ha{} emitters (xHAEs) to estimate the total \ha{} flux. We similarly construct the [NII], [OIII], [SII]$\lambda$6717, 6731 and H$\beta$ emission line maps, and measure the total flux in each of the emission lines respectively. We also estimate the total error in the emission line flux inside the aperture by adding the error in each spaxel in quadrature. We do this for each of the four emission line maps and for all the outlying \ha{} emitters in our sample. Table \ref{int_flux_ext} shows the integrated fluxes of the individual sources in our sample. We also extracted the integrated spectrum in these apertures for each source from the MaNGA datacube which is shown in Appendix \ref{integrated_spectrum} available with the online version of this paper.

\begin{table*}
\caption{Integrated fluxes and uncertainity of \ha{}, [OIII], [NII], H$\beta$, [SII]6717, and [SII]6731 emission lines for the extended outlying \ha{} emitters.}
\begin{center}
\begin{tabular}{cccccccccc}
\hline \\
ObJ ID &  F$_{\text{H}\alpha}$ ($\sigma_{\text{H}\alpha}$) & F$_{\text{[OIII]}}$ ($\sigma_{\text{[OIII]}}$) & F$_{\text{[NII]}}$ ($\sigma_{\text{[NII]}}$) & F$_{\text{H}\beta}$ ($\sigma_{\text{H}\beta}$) & F$_{\text{[SII]}{6717}}$ ($\sigma_{\text{[SII]}{6717}}$) & F$_{\text{[SII]}{6731}}$ ($\sigma_{\text{[SII]}{6731}}$) & BPT classification \\
\hline
X1 & 2.166 (0.033) & 0.884 (0.053) & 0.894 (0.038) & 0.595 (0.048) & 2.004 (0.117) & 0.302 (0.113) & composite/LINER \\
X2.1 & 2.641 (0.033) & 1.39 (0.044) & 0.804 (0.042) & 0.772 (0.037) & 0.472 (0.065) & 0.256 (0.072) & composite/star forming  \\
X2.2 & 7.323 (0.039) & 15.332 (0.098) & 2.208 (0.067) & 2.357 (0.053) & 1.338 (0.102) & 0.549 (0.092) & Seyfert  \\
X3 & 2.464 (0.037) & 0.778 (0.055) & 0.565 (0.05) & 0.594 (0.064) & 1.027 (0.034) & 0.368 (0.04) & star forming \\
X4 & 4.085 (0.051) & 9.535 (0.102) & 5.525 (0.12) & 1.603 (0.091) & 3.655 (0.16) & 1.672 (0.17) & Seyfert/LINER\\
X5 & 4.218 (0.055) & 7.1 (0.087) & 0.45 (0.065) & 1.421 (0.066) & 0.525 (0.056) & 0.403 (0.052) & star forming/Seyfert \\
\hline
\end{tabular} \label{int_flux_ext}

\end{center}
Notes .\ Column (1) gives the object ID of the outlying \ha{} emitter. Columns (2), (3), (4), (5), (6) and (7) give the total flux and 1$\sigma$ uncertainity for the \ha{}, [OIII], [NII], H$\beta$, [SII]$\lambda$ 6717 and [SII]$\lambda$ 6731  line respectively. For non-detections, the fluxes are replaced by the 3-$\sigma$ upper limits. All values are in units of $10^{-16}$ergs s$^{-1}$ cm$^{-2}$. Column (8) gives the BPT classification of the xHAEs. See Sec. \ref{BPT} for details.  

\end{table*}

\subsubsection{BPT diagram}\label{BPT}
\begin{figure*}
\begin{center}
\includegraphics[scale=0.65]{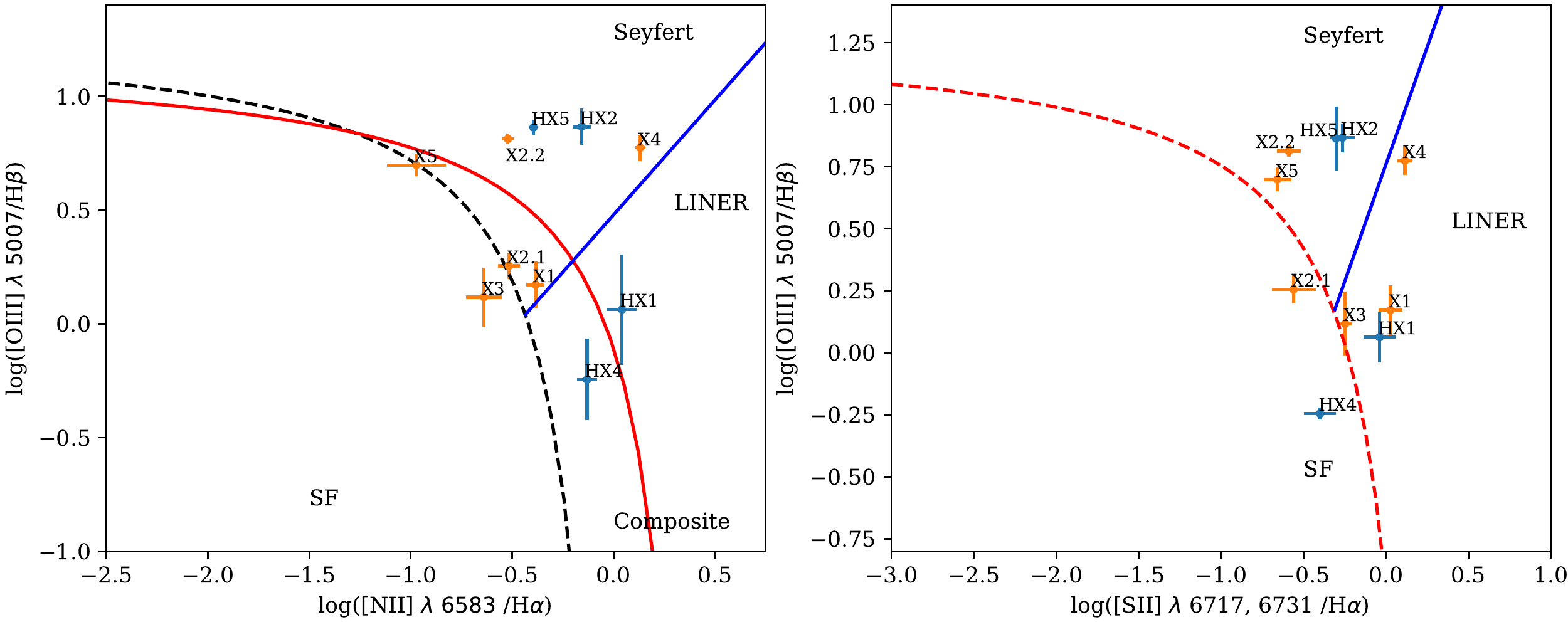}

\caption{ Left Panel: The [OIII]/H$\beta$ vs. [NII]/\ha{} BPT diagnostic diagram. The red solid line is the extreme starburst line from \citet[][Ke06]{Kewley06}. The black dashed line is from the semi-empirical fit to the SDSS galaxies from \citet[][Ka03]{Kauffmann03AGN}. The region below the Ka03 line is termed as star-forming (SF) region. In the region above the Ke01, the blue solid line from \citet{CidFernandes10} seperates the Seyferts from LINERS. The region between the Ke06 and Ka03 line is termed as composite region, which has emission from both star formation and AGN.  Right Panel: The [OIII]/H$\beta$ vs. [SII]$\lambda$ 6717, 6731/\ha{} BPT diagnostic diagram. The red dashed line is the extreme starburst line from \citet[][Ke06]{Kewley06}. The extended \ha{} emitters are shown in orange dots. These show various sources of ionisation in their spectrum. The blue points are for the corresponding host galaxy.  See the text for details.} \label{BPT_plot}

\end{center}
\end{figure*}

In this section, we use the BPT diagnostic diagram to identify the source of ionisation in the outlying \ha{} emitters. The BPT diagram is widely used to classify galaxies into star forming (SF), Seyferts, and low-ionisation emission line regions (LINERs) \citep[e.g.,][]{Kewley06, Kauffmann03AGN}. The BPT classification scheme is also used in resolved IFU studies \citep[e.g.,][]{Belfiore16, Lin17}. In our study we use the [OIII]$\lambda$5007/H$\beta$ vs. [NII]$\lambda$6583 ([SII]$\lambda$6717, 6731)/\ha{} diagram. For simplicity, we will refer to the [OIII]$\lambda$5007, [NII]$\lambda$6583, [SII]$\lambda$6717, 6731 (which is the sum of the two lines) as [OIII], [NII], and [SII] respectively.

Figure \ref{BPT_plot} left (right) panel shows the [OIII]/H$\beta$ vs. [NII]([SII])/\ha{} diagnostic diagram for our sample of outlying \ha{} emitters. The corresponding line ratios for the host galaxies are determined using SDSS DR14 spectroscopy. In both the panels the red solid line is  for the theoretical maximum starburst line from \citet[][]{Kewley06} which we refer as the Ke06 line hereafter. And the black dashed line in the left panel is semi-empirical line from \citet[][]{Kauffmann03AGN} which we refer to as the Ka03 line hereafter. In the [NII]/H$\beta$ BPT diagram the source of ionisation for the region below the Ka03 line is purely from star formation. For the region above the Ke01 line the source of ionisation is from a type 2 AGN wherein it is from a Seyfert AGN if it is above the blue  solid line and LINER if it is below the blue solid line \citep{CidFernandes10}. For the region between the Ka03 and Ke01 lines the source of ionisation is from  a mix of star formation and non-star formation, and it is termed as ``composite". Moreover, ionisation due to slow shocks has been also shown to push the points into the LINER region \citep{Farage10, Rich10}. In the [SII]/\ha{} BPT diagram, the region below the Ke06 line has ionisation purely due to star formation. The region above the Ke06 line and above the blue line has ionisation due to Seyfert AGN and region above the Ke06 line and below the blue line has ionisation due to LINER like AGN.

In Figure \ref{BPT_plot}, the most striking result is that of X2.2, which is classified as ionized due to a Seyfert AGN in both the diagrams. X2.1, which is at the edge of extreme starburst line in the [NII]/\ha{} BPT diagram, has moved within the starburst line in the [SII]/\ha{} diagram. Interestingly, the corresponding host galaxy (HX2) has a Seyfert AGN in the centre, which provides the ionising photons for X2.2, thus putting it in the Seyfert region. X1 is classified as composite and LINER in the [NII]/\ha{} and the [SII]/\ha{} BPT diagrams respectively. The host (HX1) also shows a corresponding LINER emission line ratio in its centre.  X4 is classified as Seyfert and LINER in the [NII]/\ha{} and the [SII]/\ha{} BPT diagram respectively. However, peculiarly the host (HX4) does not show any nuclear activity. X5 which is classified as at the edge of extreme starburst in the [NII]/\ha{} diagram is shifted to the Seyfert region in the [SII]/\ha{} diagram. The host galaxy (HX5) also shows a Seyfert-like line ratios in the centre, and thus could be the source of ionisation. X3 is still consistent with being at the edge of extreme starburst line from both the BPT diagrams. The host galaxy (HX3) is devoid of any emission lines and thus does not have any nuclear activity or SF in the centre (and hence it is not shown in Figure \ref{BPT_plot}). 

Thus  the extended emitters show a mix sources of ionisation ranging form star formation to composite emission. In particular X2.2 and X4 show clear signatures of Seyfert/LINER like emission line ratios. Except in the case of X4, the corresponding host galaxies have a active nuclei which could be the source of ionisation of these regions. The extended morphology and Seyfert/LINER like emission line ratios lead us to think that these objects are extended emission line regions (EELRs) e.g. the Hanny's Voorwerp \citep{Lintott09, Keel12a} and similar EELRs searched by the Galaxy Zoo candidates \citep{Keel12b},  and \citet{Unger87} and \citet{Tadhunter89}. We describe each of these objects in detail in the next section.
\subsection{Comparison of \ha{} luminosities of xHAEs with literature}
\label{comparison}

\citet{Keel12b}, using the Galaxy Zoo candidates, have carried out a systematic search for extended emission line regions (EELRs) and have documented 19 objects. A detailed study of the sources of ionisation in their EELRs shows that they are  photoionised due to an AGN.  The objects reported in our sample also show similar morphological features as those in the \citet{Keel12b} sample and are also primarily photoionised due to an AGN. However, unlike their sample, our sample objects do not show any bright optical counterparts in deep optical images. The median values of the \ha{} luminosities of our xHAEs is $4.47\times 10^{39}$ergs/s. This is about 4.5 times fainter than the median \ha{} luminosity of the \citet{Keel12b} sample ($2\times 10^{40}$ ergs/s). Thus, using our selection method with MaNGA data we are able to detect fainter counterparts of Hanny's Voorwerp like objects. 

In conclusion, we find that the xHAEs have ionisation due to various kinds of sources ranging from Seyfert AGN, star formation and also a mixture of both. Moreover, their \ha{} luminosities show that they are fainter counterparts of the EELRs.

\section{Remarks on individual sources}\label{notes}
In this section we will briefly describe the xHAEs in our sample .

\begin{figure*}

\includegraphics[scale=0.5]{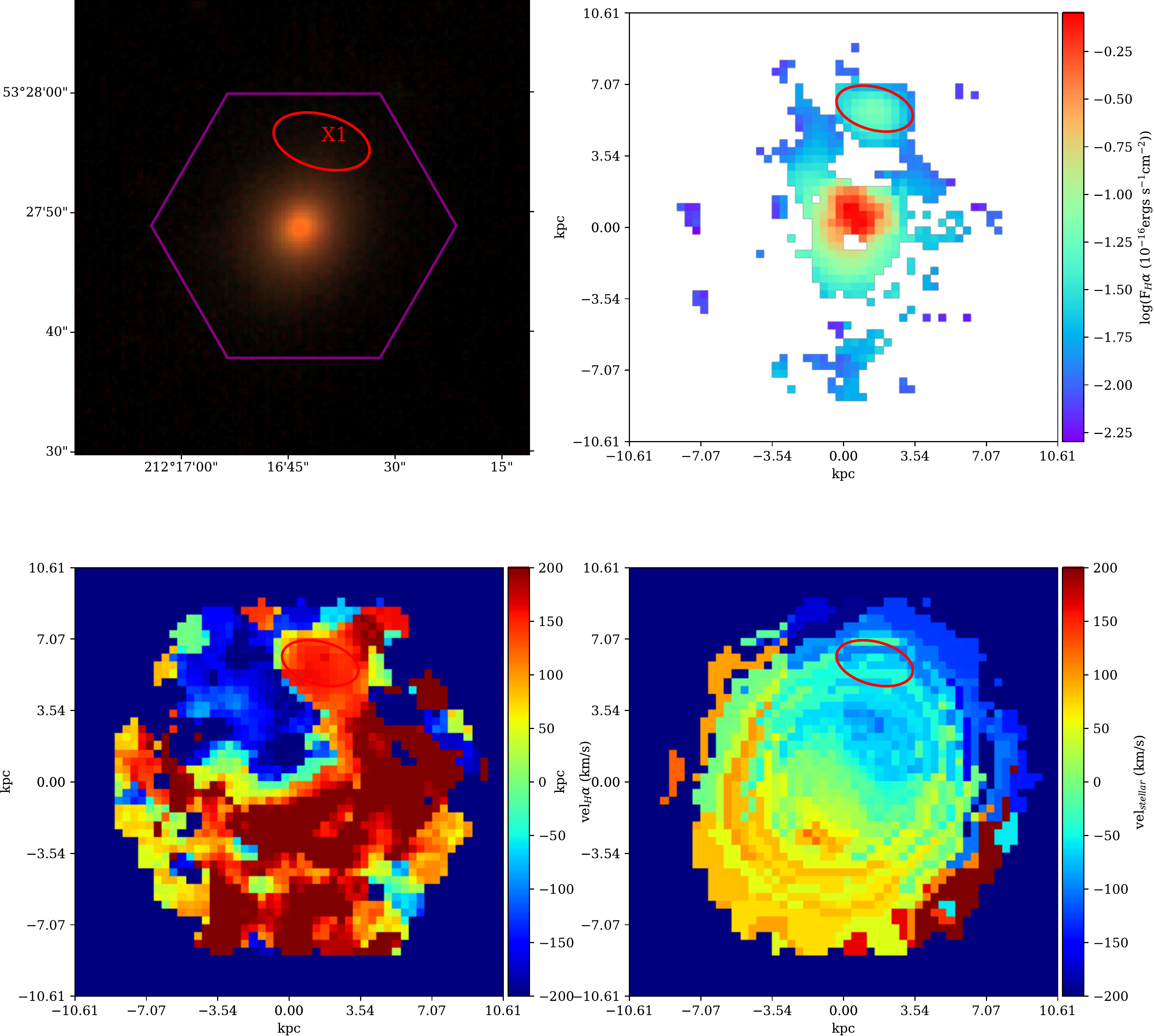}

\caption{Top-left panel: Shows the DECaLs or MzLS-BASS $grz$  \citet{Lupton04} colour composite image of the host galaxies with the MaNGA IFU extent in purple. Top-right panel, bottom-left and bottom-right panel shows the \ha{} emission map, \ha{} velocity map and stellar velocity map respectively for X1 in physical units in kpcs. In each panel the red ellipse highlights the location of the xHAEs. See text for details.} \label{montage_1}

\end{figure*}

\begin{figure*}

\includegraphics[scale=0.5]{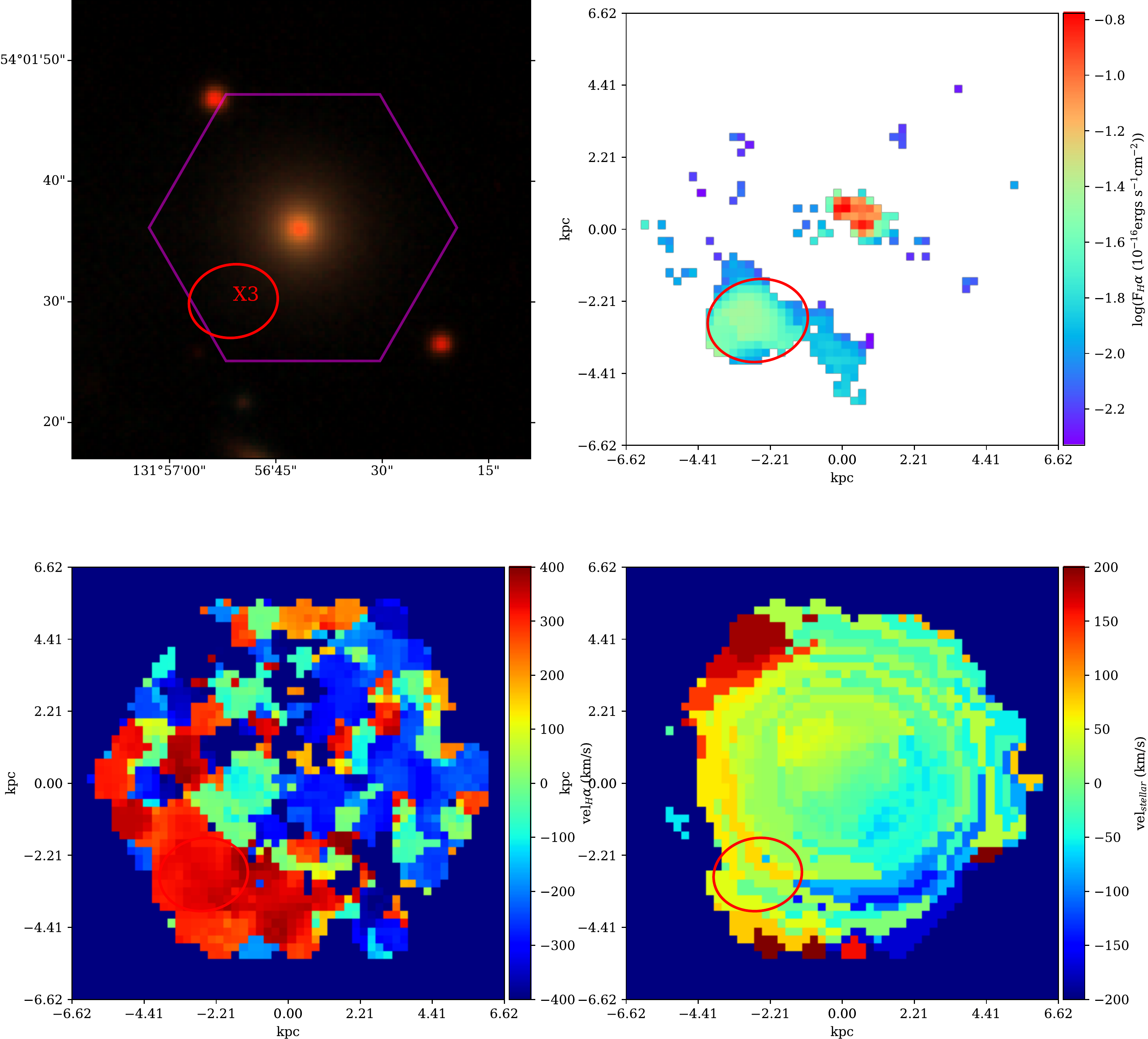}

\caption{Continued. Here we show the outlying emission for X3.} \label{montage_3}

\end{figure*}

\begin{figure*}

\includegraphics[scale=0.5]{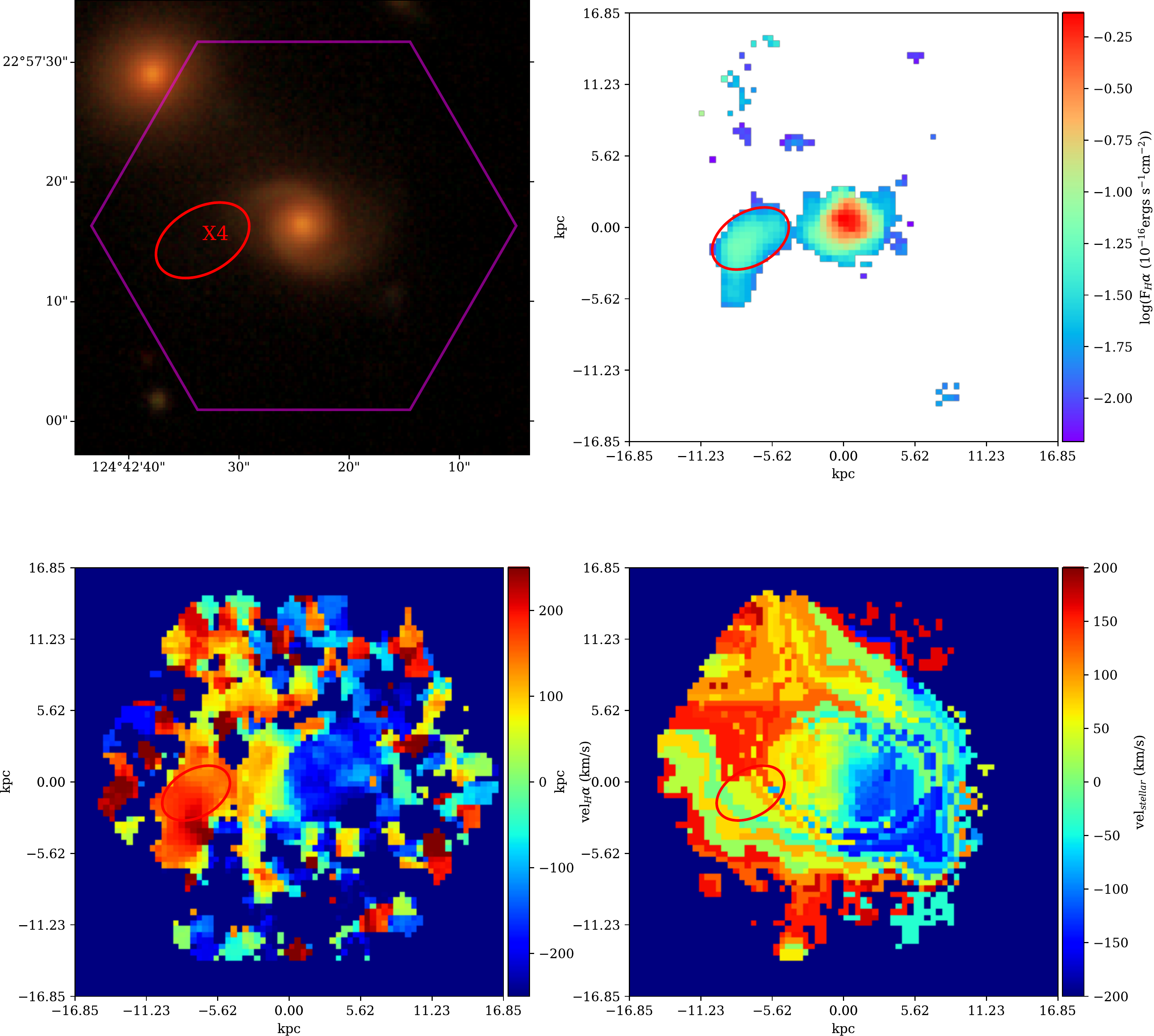}

\caption{Continued. Here we show the outlying emission for X4.} \label{montage_4}

\end{figure*}

\begin{figure*}

\includegraphics[scale=0.5]{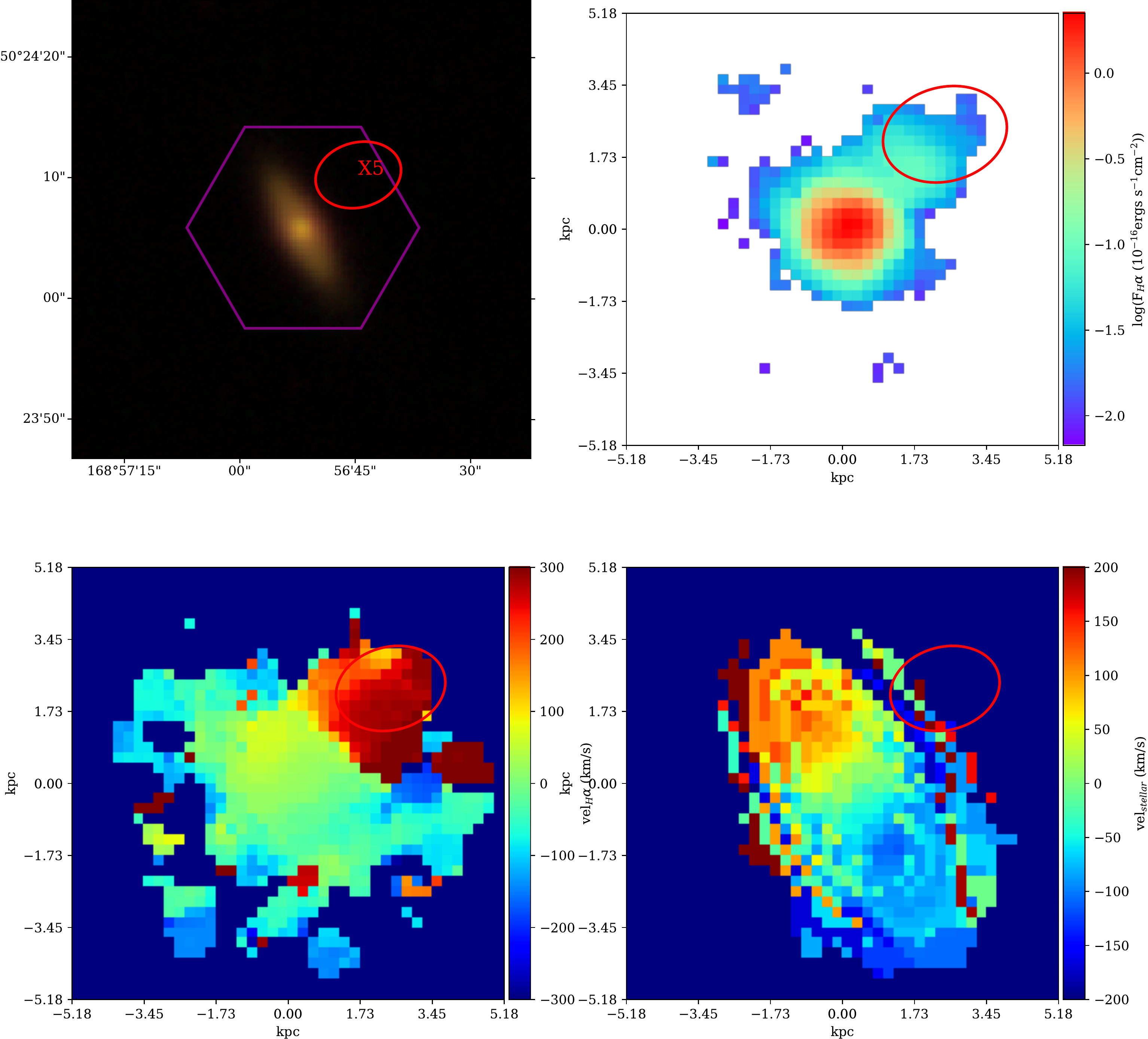}

\caption{Continued. Here we show the outlying emission for X5.} \label{montage_5}
\end{figure*}

\begin{enumerate}
\item X1 (host SDSS objID: 1237661416601878596): As shown in the \ha{} image in Figure \ref{montage_1}, there is an extended outlying \ha{} emission (highlighted with a red ellipse and referred to as X1). The host galaxy, which has an elliptical morphology \citep[][G18 hereafter]{Graham18}, does not show any features of interaction in its light profile both at the location of the \ha{} emission and elsewhere. The \ha{} velocity map  of the galaxy on the contrary shows very disturbed features. There appears to be some kind of rotation along the northeast-southwest diagonal axis, with the gas in the northeast direction moving away from us and the gas along the southwest direction coming towards us. The outlying \ha{} emission on the other hand has a different velocity from this, and is coming towards us with an velocity of about 100 km/s. Moreover, the stellar velocity shown in the last panel is completely misaligned from the velocity flow of the \ha{} emission. Such an misalignment is a signature of recent gas rich merger. X1 has a composite/LINER emission line ratios from the BPT diagram (see Table \ref{int_flux_ext}), and the host galaxy is classified has LINER like emission line ratios in the centre. This could be an example of Hanny's Voorwerp like object. However, unlike Hanny's Voorwerp X1 has no detectable optical counterpart and hence this object is likely a fainter counterpart of the Voorwerp. 

\item X2.1 \& X2.2 (host SDSS objID: 1237655374109802660): In Fig. \ref{example}, both X2.1  and X2.2  have emission coming from a merging system of two galaxies (G18). Here we identify the host galaxy not with the galaxy on which the MaNGA IFU is centered (ObJID: 1237655374109802660), but rather with the galaxy adjacent to it (ObJID: 1237655374109802661), due to the continuity of the \ha{} emission with it. X2.1 has a velocity very near the velocity of the host galaxy and could be a case of extraplanar gas. This merger is caught in the MaNGA IFU field of view. The IFU is centred around the galaxy 8601-12701. X2.1 is at the edge of composite/star forming region, and X2.2 has Seyfert emission line ratios. The host galaxy (HX2) also has Seyfert AGN in the centre. It is intriguing to see that both X2.1 and X2.2 emitting regions which are on opposite sides of the host galaxy show a connected bridge. Moreover, the velocity map shows a gradient as we go along the connecting bridge around the host galaxy. Both X2.1 and X2.2 are examples of fainter counterparts of Hanny's Voorwerp. 


\item X3 (host SDSS objID: 1237651533336281291, see Figure \ref{montage_3}): The host galaxy has an elliptical morphology (G18) with no clear indication of recent merger/integration in its light profile. The host galaxy of X3 is however, devoid of any emission lines and thus does not have an active nucleus. Interestingly, X3 is a large star forming region, however it has a velocity which is very different (about 400 km/s) from the disc as seen in the stellar velocity profile (due to poor \ha{} SNR of the host galaxy the velocity map is not visible). It is possible that this galaxy may have undergone a recent merger with a galaxy which may have left a tidal tail or there has been a recent event of gas accretion as in the case of \citet{Cheung16b}. Given that there is such a high velocity difference, it is also possible that the \ha{} emission is not associated with HX3, and is rather associated with a neighbouring galaxy (SDSS ObjID: 1237651252560461948), which is about 570 km/s offset from it, which also has an active nucleus. Thus, a past interaction with HX3 may have left a tidal tail, which is being photoionised by the neighbouring galaxy. 

\item X4 (host SDSS objID: 1237661085345513774, see Figure \ref{montage_4}): The host galaxy is a spiral galaxy with a close companion nearby (G18) and has a LINER/composite emission line ratios. It is possibly interacting with another early type galaxy. X4 has Seyfert/LINER like emission line ratio. The \ha{} gas and stellar rotation is in the same direction although there is an misalignment between the two. X4 is also an example of Hanny's Voorwerp like object.    

\item X5 (host SDSS objID: 1237657855533056062, see Figure \ref{montage_5}): X5, like X2.1 and X2.2, also shows a continuity with the host galaxy. The host of X5 is an S0 galaxy (G18) which has a Seyfert AGN in the centre. Interestingly, X5 is at the edge of the star forming/Seyfert region. The \ha{} velocity map shows that the gas has very different velocity from the host galaxy and is moving away from the galaxy with a velocity of at least 300 km/s, which is fairly large. However, given the continuity in the \ha{} emission with the host galaxy we believe that it is associated with it. It is possible that this is a event of outflowing gas due to AGN feedback, and this gas is undergoing in-situ star formation. \citet{Maiolino17} have recently observed such an event of in-situ star formation in galactic outflows at a redshift of 0.0448. Intriguingly, the \ha{} velocity map shows a gradient, and thus there are likely multiple velocity components in the emission region. This feature is also seen in the integrated spectrum shown in the Appendix \ref{integrated_spectrum} (Figure A6), where we see two velocity components in the spectrum.  

\end{enumerate}

\section{Summary}\label{conclusion}
In summary, we have conducted a systematic search for outlying \ha{} emitters in the entire \piped{} value added catalogue of the recently released DR14 SDSS IV MaNGA survey of 2,755 galaxies. We have found six outlying  extended \ha{} emitters (xHAEs) which have strong \ha{} emission away from the host galaxy and without any bright optical continuum counterpart in the deeper optical images from the DECaLS or MzLS/BASS survey. These \ha{} emitters also show a \ha{} velocity component which is different from that of the host galaxy, ranging from about 150-400 km/s. We have also found several artefacts in the MaNGA datacubes most of which appear unresolved sources in the \ha{} maps. However, their \ha{} line-widths are much narrower than the instrumental width of the MaNGA IFU. These are likely to be spurious cosmic rays which were not flagged by the MaNGA data reduction pipeline. The integrated line flux ratios show  of the six xHAEs shows various kinds of sources of photoionisation. Three of the xHAEs show photoionisation due to an AGN, two of them have composite emission line ratio and one of them have photoionisation due to star formation. These xHAEs have \ha{} luminosities at the lower end of the previously known EELRs and are thus their fainter counterparts. 

Due to the poor spatial resolution of the MaNGA IFU it is very difficult to study the emission line regions with more detail, in particular since it can blend emission lines from two different, but spatially nearby, sources of ionisation into a single line. To address this limitation, we intend to follow-up these objects in our sample with a higher spatial resolution IFU e.g., SITELLE
(Spectrom{\`e}tre Imageur {\`a} Transform{\'e}e de Fourier pour l'Etude en Long et en Large de raies d'Emission) on the Canada-France-Hawaii-Telescope (CFHT). This will also give us a larger field of view which can capture more emission line regions which are missed in the narrow field of view of MaNGA. Along with these new IFU observations, we also hope to obtain deep optical images to locate the optical counterparts of the \ha{} emitters in our sample.

\section*{Acknowledgments}

We thank the anonymous referee for insightful comments that have
improved both the content and presentation of this paper. We would also like to thank Alexei Moiseev for alerting us to artefacts in the MaNGA datacubes.

Funding for the Sloan Digital Sky Survey IV has been provided by the Alfred P. Sloan Foundation, the U.S. Department of Energy Office of Science, and the Participating Institutions. SDSS acknowledges support and resources from the Center for High-Performance Computing at the University of Utah. The SDSS web site is www.sdss.org.

SDSS is managed by the Astrophysical Research Consortium for the Participating Institutions of the SDSS Collaboration including the Brazilian Participation Group, the Carnegie Institution for Science, Carnegie Mellon University, the Chilean Participation Group, the French Participation Group, Harvard-Smithsonian Center for Astrophysics, Instituto de Astrofísica de Canarias, The Johns Hopkins University, Kavli Institute for the Physics and Mathematics of the Universe (IPMU) / University of Tokyo, Lawrence Berkeley National Laboratory, Leibniz Institut für Astrophysik Potsdam (AIP), Max-Planck-Institut für Astronomie (MPIA Heidelberg), Max-Planck-Institut für Astrophysik (MPA Garching), Max-Planck-Institut für Extraterrestrische Physik (MPE), National Astronomical Observatories of China, New Mexico State University, New York University, University of Notre Dame, Observatório Nacional / MCTI, The Ohio State University, Pennsylvania State University, Shanghai Astronomical Observatory, United Kingdom Participation Group, Universidad Nacional Autónoma de México, University of Arizona, University of Colorado Boulder, University of Oxford, University of Portsmouth, University of Utah, University of Virginia, University of Washington, University of Wisconsin, Vanderbilt University, and Yale University.

This project makes use of the MaNGA-Pipe3D dataproducts. We thank the IA-UNAM MaNGA team for creating this catalogue, and the ConaCyt-180125 project for supporting them.

The Legacy Surveys consist of three individual and complementary projects: the Dark Energy Camera Legacy Survey (DECaLS; NOAO Proposal ID \# 2014B-0404; PIs: David Schlegel and Arjun Dey), the Beijing-Arizona Sky Survey (BASS; NOAO Proposal ID \# 2015A-0801; PIs: Zhou Xu and Xiaohui Fan), and the Mayall z-band Legacy Survey (MzLS; NOAO Proposal ID \# 2016A-0453; PI: Arjun Dey). DECaLS, BASS and MzLS together include data obtained, respectively, at the Blanco telescope, Cerro Tololo Inter-American Observatory, National Optical Astronomy Observatory (NOAO); the Bok telescope, Steward Observatory, University of Arizona; and the Mayall telescope, Kitt Peak National Observatory, NOAO. The Legacy Surveys project is honored to be permitted to conduct astronomical research on Iolkam Du'ag (Kitt Peak), a mountain with particular significance to the Tohono O'odham Nation.

NOAO is operated by the Association of Universities for Research in Astronomy (AURA) under a cooperative agreement with the National Science Foundation.

This project used data obtained with the Dark Energy Camera (DECam), which was constructed by the Dark Energy Survey (DES) collaboration. Funding for the DES Projects has been provided by the U.S. Department of Energy, the U.S. National Science Foundation, the Ministry of Science and Education of Spain, the Science and Technology Facilities Council of the United Kingdom, the Higher Education Funding Council for England, the National Center for Supercomputing Applications at the University of Illinois at Urbana-Champaign, the Kavli Institute of Cosmological Physics at the University of Chicago, Center for Cosmology and Astro-Particle Physics at the Ohio State University, the Mitchell Institute for Fundamental Physics and Astronomy at Texas A\&M University, Financiadora de Estudos e Projetos, Fundacao Carlos Chagas Filho de Amparo, Financiadora de Estudos e Projetos, Fundacao Carlos Chagas Filho de Amparo a Pesquisa do Estado do Rio de Janeiro, Conselho Nacional de Desenvolvimento Cientifico e Tecnologico and the Ministerio da Ciencia, Tecnologia e Inovacao, the Deutsche Forschungsgemeinschaft and the Collaborating Institutions in the Dark Energy Survey. The Collaborating Institutions are Argonne National Laboratory, the University of California at Santa Cruz, the University of Cambridge, Centro de Investigaciones Energeticas, Medioambientales y Tecnologicas-Madrid, the University of Chicago, University College London, the DES-Brazil Consortium, the University of Edinburgh, the Eidgenossische Technische Hochschule (ETH) Zurich, Fermi National Accelerator Laboratory, the University of Illinois at Urbana-Champaign, the Institut de Ciencies de l\' Espai (IEEC/ CSIC), the Institut de Fisica d\' Altes Energies, Lawrence Berkeley National Laboratory, the Ludwig-Maximilians Universitat Munchen and the associated Excellence Cluster Universe, the University of Michigan, the National Optical Astronomy Observatory, the University of Nottingham, the Ohio State University, the University of Pennsylvania, the University of Portsmouth, SLAC National Accelerator Laboratory, Stanford University, the University of Sussex, and Texas A\&M University.

BASS is a key project of the Telescope Access Program (TAP), which has been funded by the National Astronomical Observatories of China, the Chinese Academy of Sciences (the Strategic Priority Research Program "The Emergence of Cosmological Structures" Grant \# XDB09000000), and the Special Fund for Astronomy from the Ministry of Finance. The BASS is also supported by the External Cooperation Program of Chinese Academy of Sciences (Grant \# 114A11KYSB20160057), and Chinese National Natural Science Foundation (Grant \# 11433005).

The Legacy Survey team makes use of data products from the Near-Earth Object Wide-field Infrared Survey Explorer (NEOWISE), which is a project of the Jet Propulsion Laboratory/California Institute of Technology. NEOWISE is funded by the National Aeronautics and Space Administration.

The Legacy Surveys imaging of the DESI footprint is supported by the Director, Office of Science, Office of High Energy Physics of the U.S. Department of Energy under Contract No. DE-AC02-05CH1123, by the National Energy Research Scientific Computing Center, a DOE Office of Science User Facility under the same contract; and by the U.S. National Science Foundation, Division of Astronomical Sciences under Contract No. AST-0950945 to NOAO.

\bibliographystyle{mnras}
\bibliography{references}

\appendix
\section{Integrated Spectrum of the extended emitters} \label{integrated_spectrum}

We used elliptical apertures around the extended emitters to extract the integrated spectrum for each of the sources in our sample, directly from the MaNGA datacube. The integrated spectra are shown in Figure \ref{int_spec_1}-\ref{int_spec_5}. We also show the host galaxy spectrum from the SDSS DR14 spectroscopic survey for reference in the bottom-most panel of each of figure.

\begin{figure*}
\begin{center}
\includegraphics[scale=0.5]{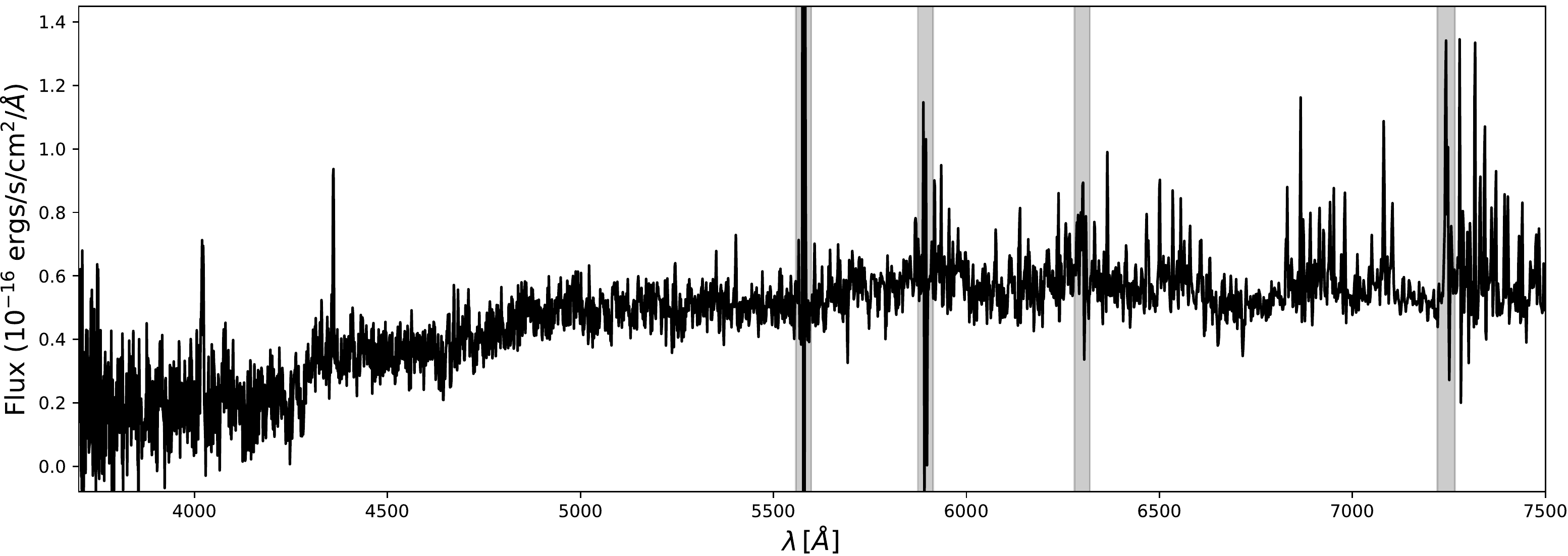}
\includegraphics[scale=0.5]{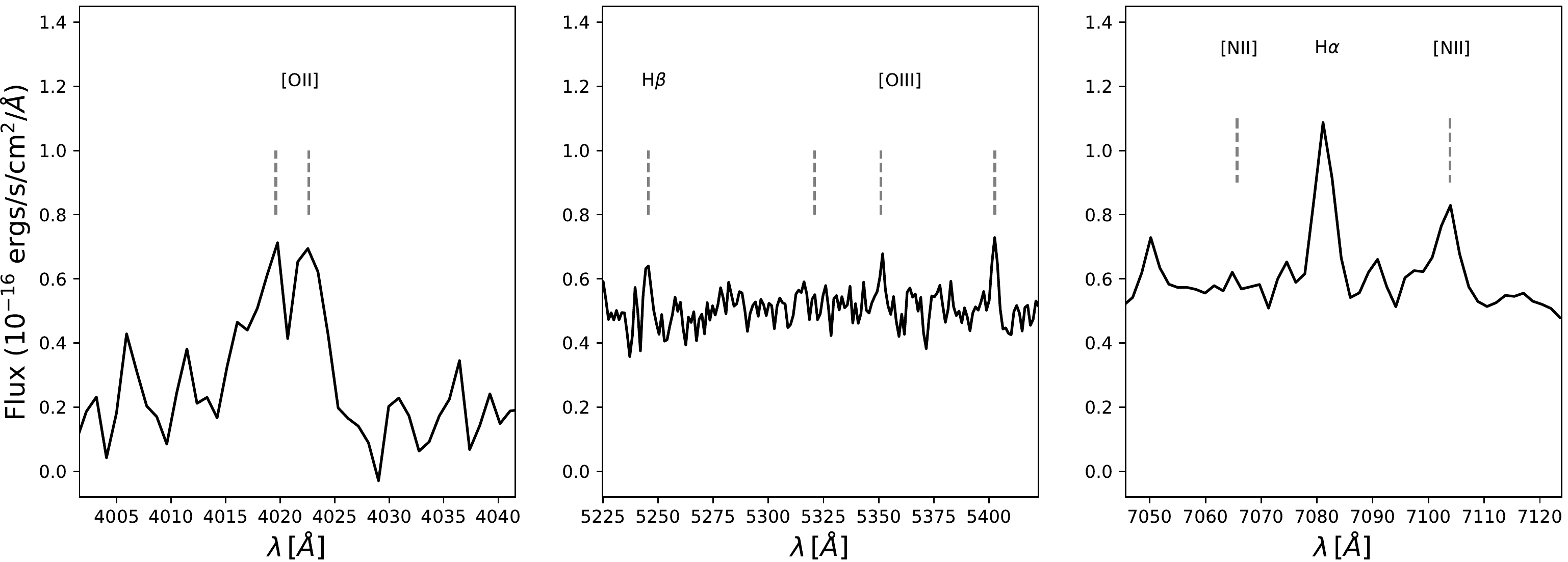}
\includegraphics[scale=0.5]{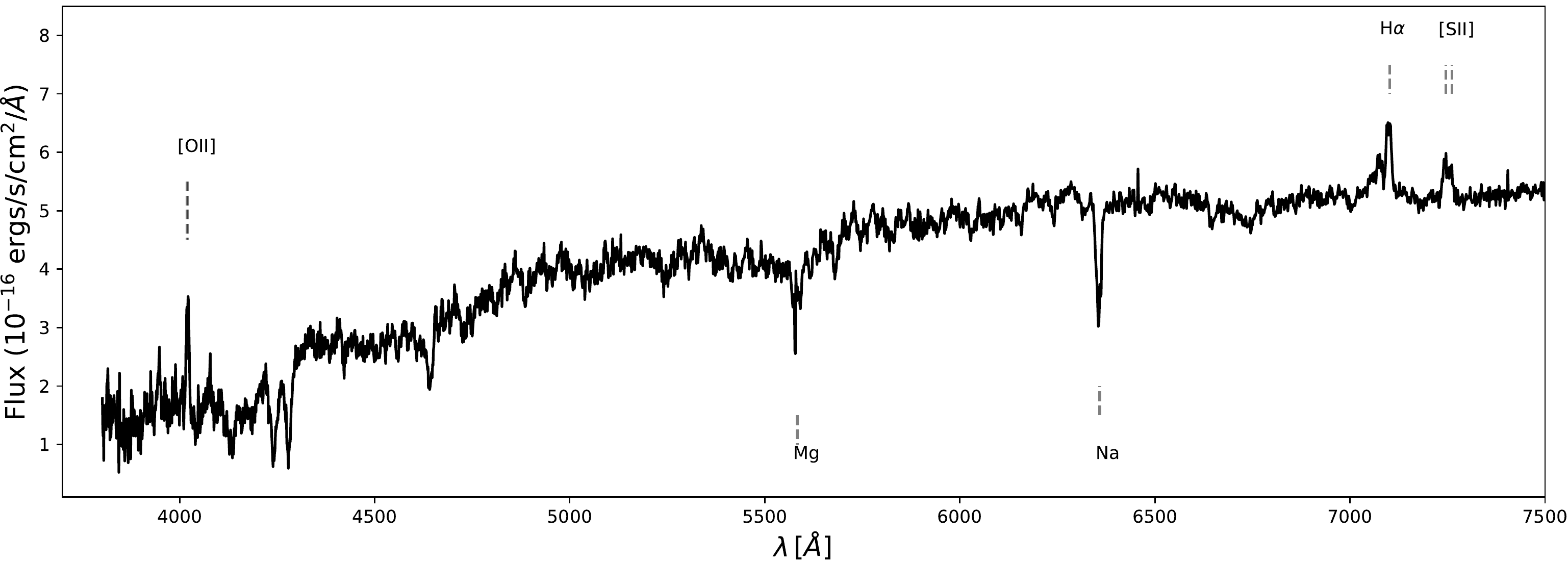}
\caption{Top: Integrated spectrum for X1. The grey shaded regions indicate regions with strong sky lines in the SDSS spectrum. Middle: A zoomed-in spectrum around the [OII] doublet (left), [OIII] $\lambda$4960, 4963, and 5007 and H$\beta$ (middle), and \ha{} and the [NII] doublet (right). Bottom: Observed frame host galaxy spectrum from the SDSS DR14 spectroscopic survey.} \label{int_spec_1}
\end{center}
\end{figure*}

\begin{figure*}
\begin{center}
\includegraphics[scale=0.5]{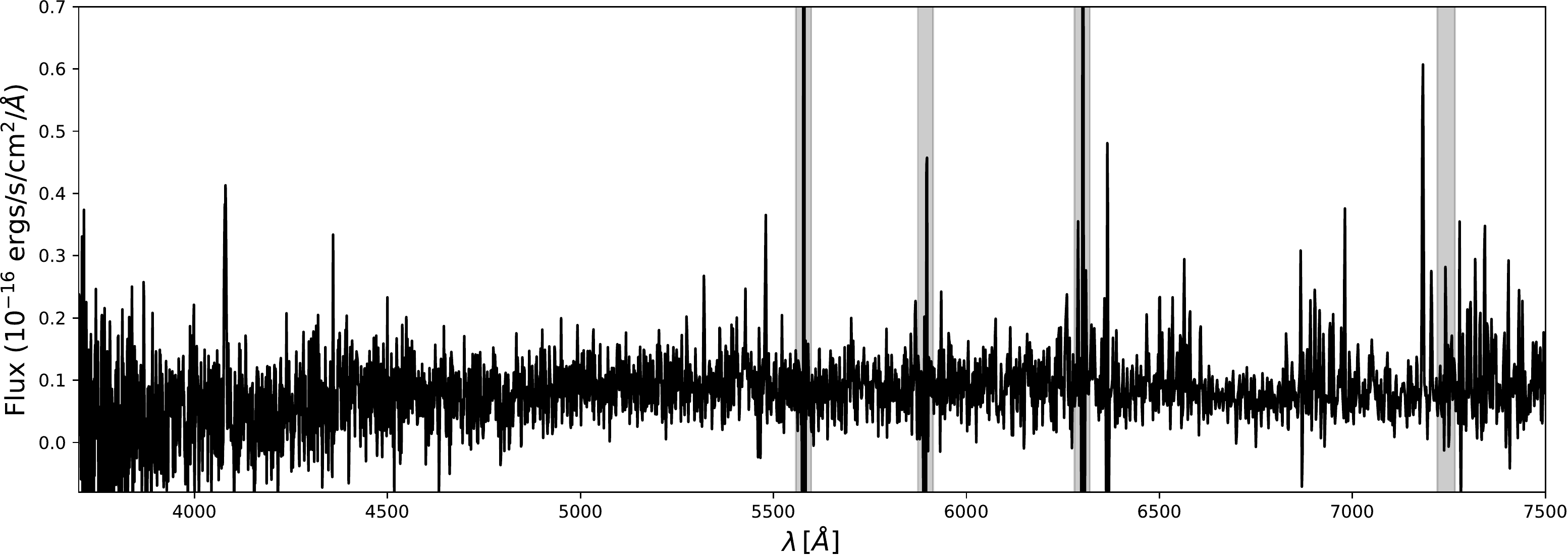}
\includegraphics[scale=0.5]{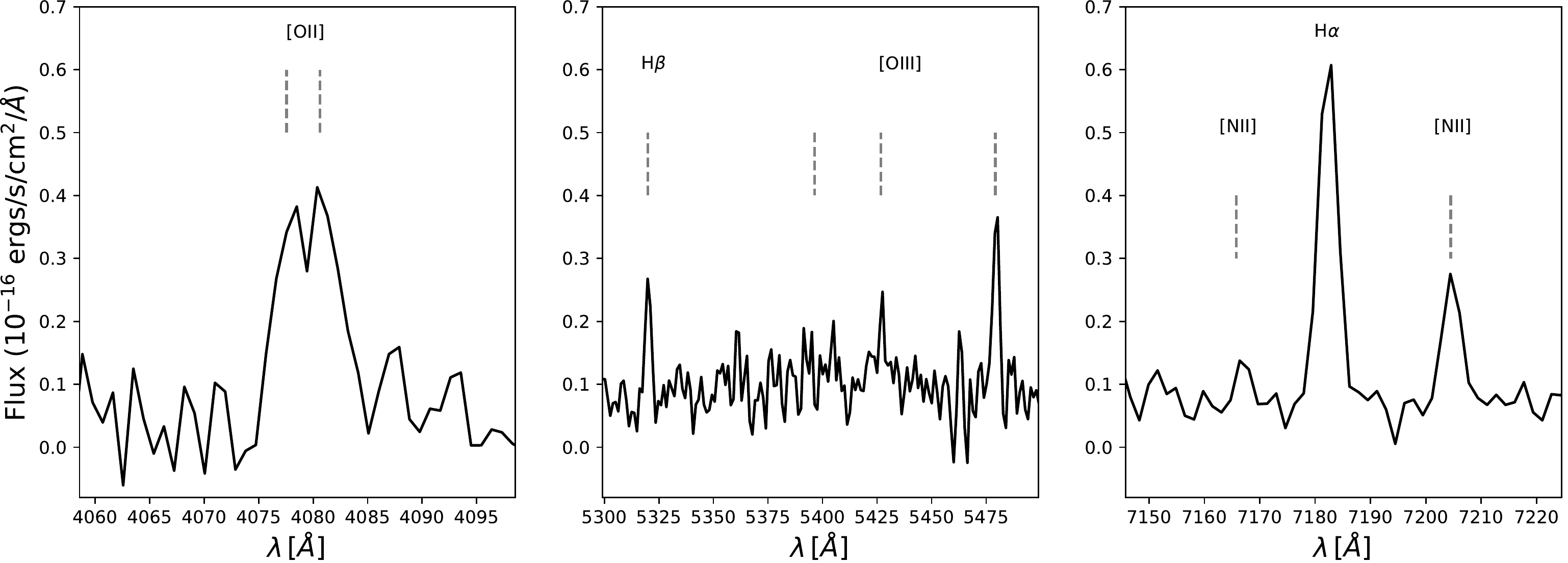}
\includegraphics[scale=0.5]{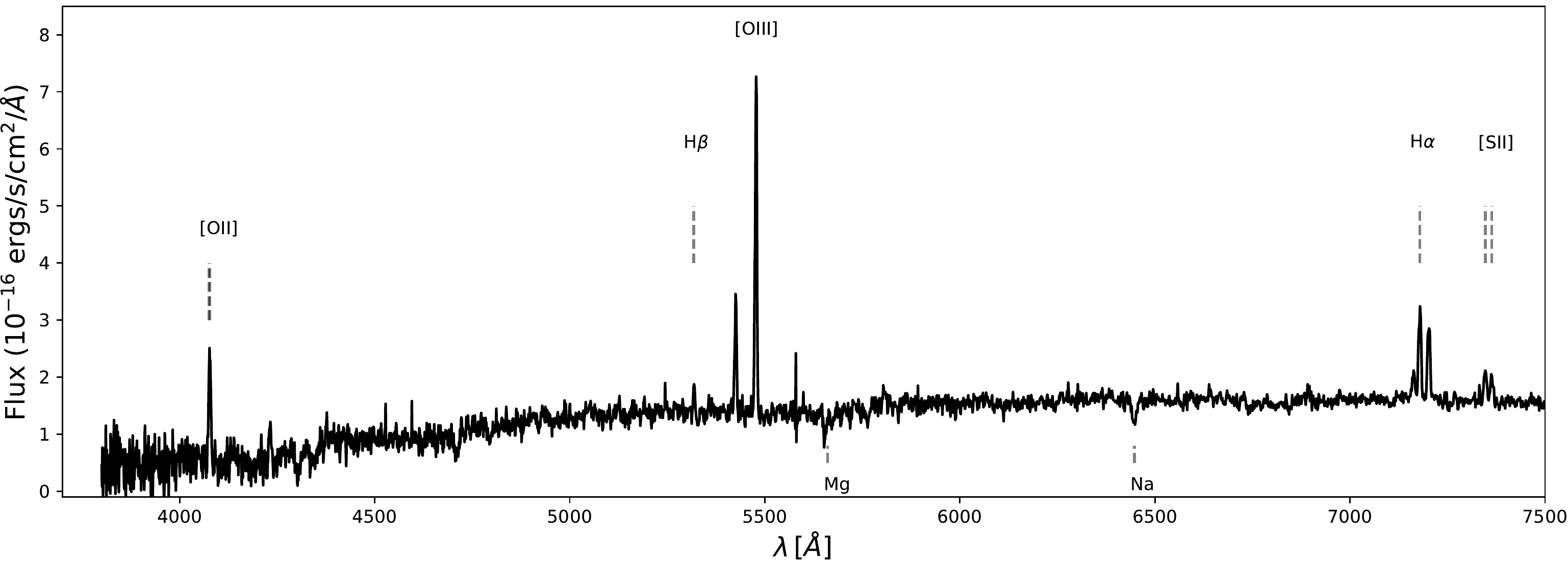}

\caption{Continued. Here we show the integrated spectrum for X2.1 and its corresponding host galaxy.}
\end{center}
\end{figure*}

\begin{figure*}
\begin{center}
\includegraphics[scale=0.5]{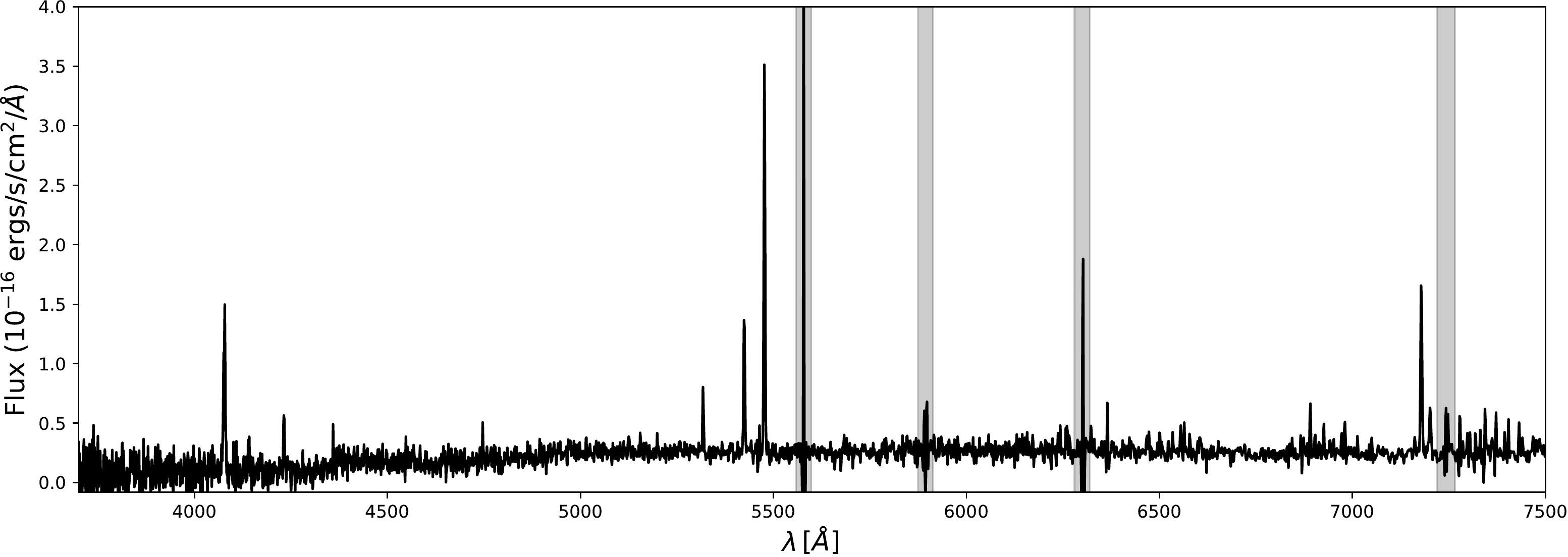}
\includegraphics[scale=0.5]{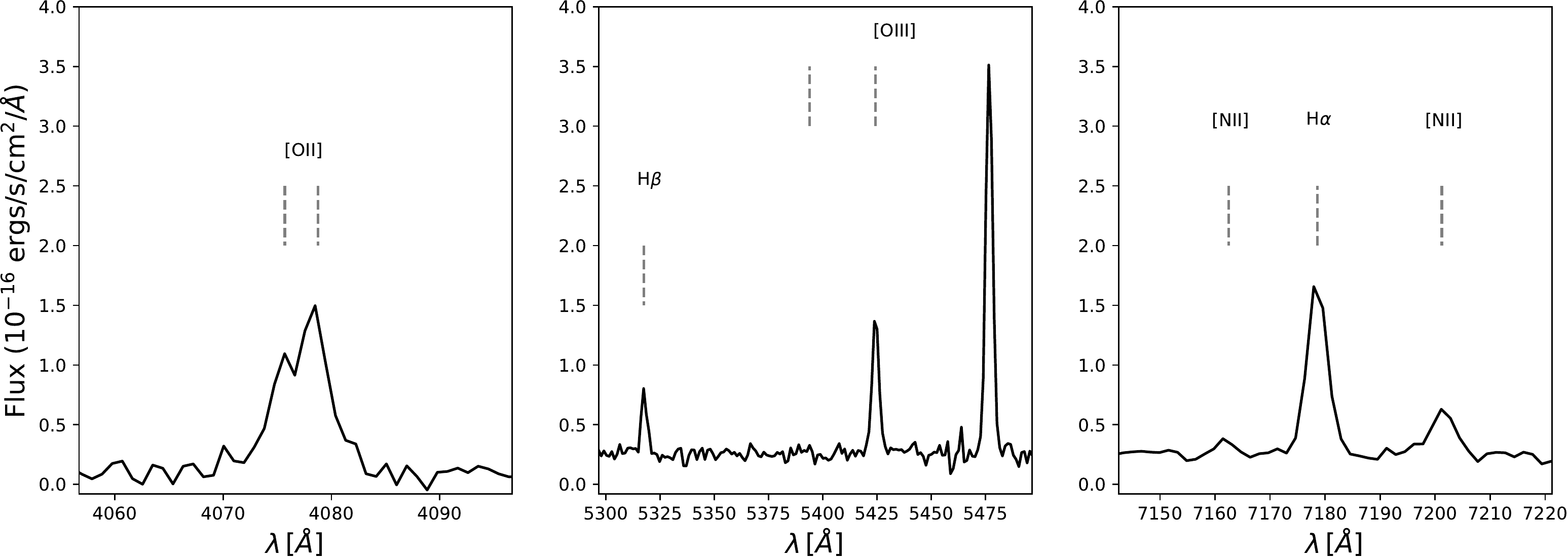}
\includegraphics[scale=0.5]{X2_host.pdf}

\caption{Continued. Here we show the integrated spectrum for X2.2 and its corresponding host galaxy.}
\end{center}
\end{figure*}

\begin{figure*}
\begin{center}
\includegraphics[scale=0.5]{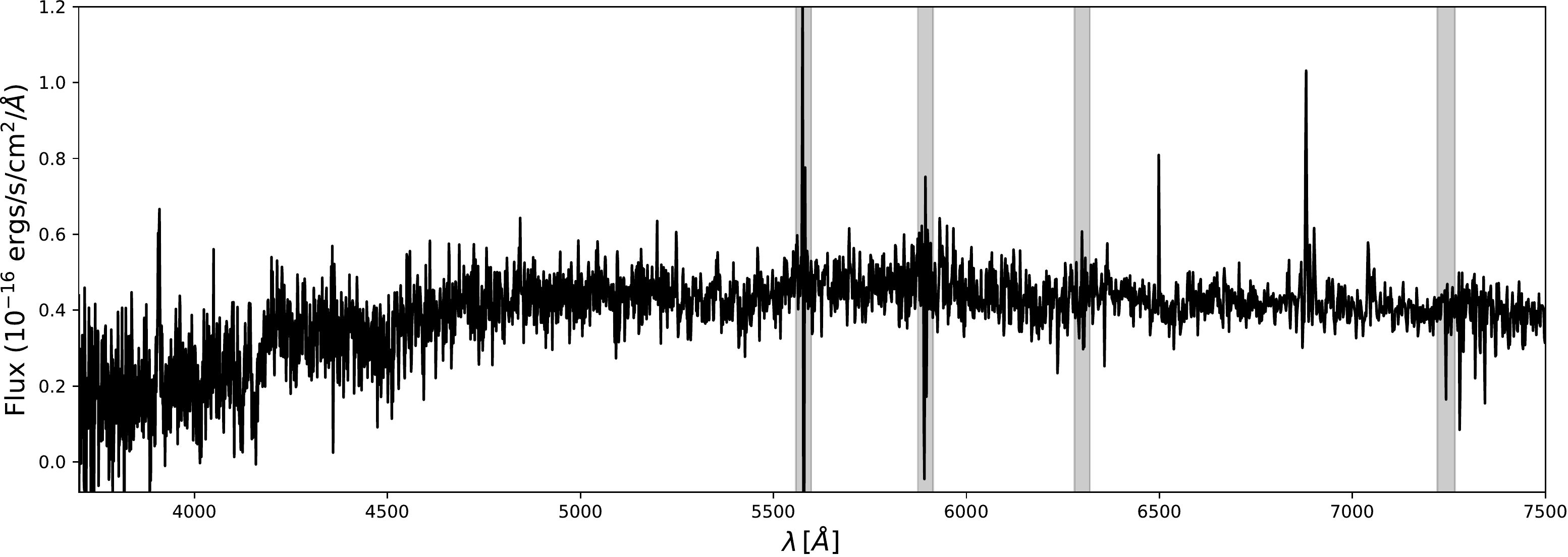}
\includegraphics[scale=0.5]{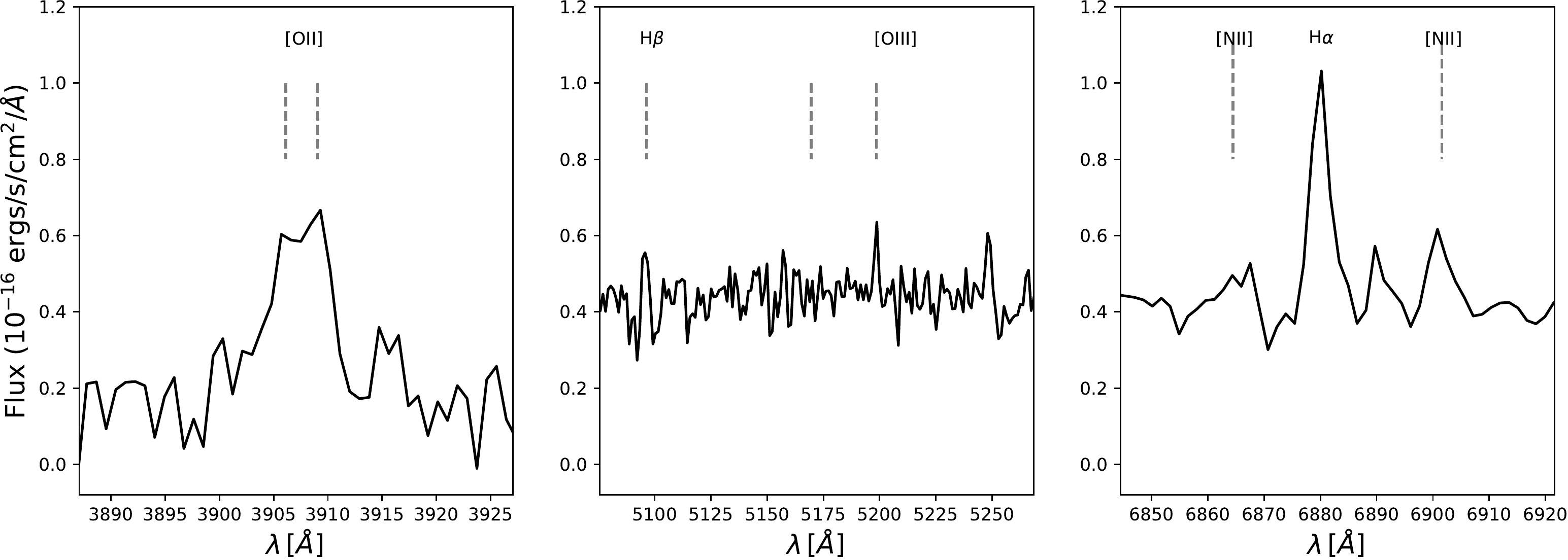}
\includegraphics[scale=0.5]{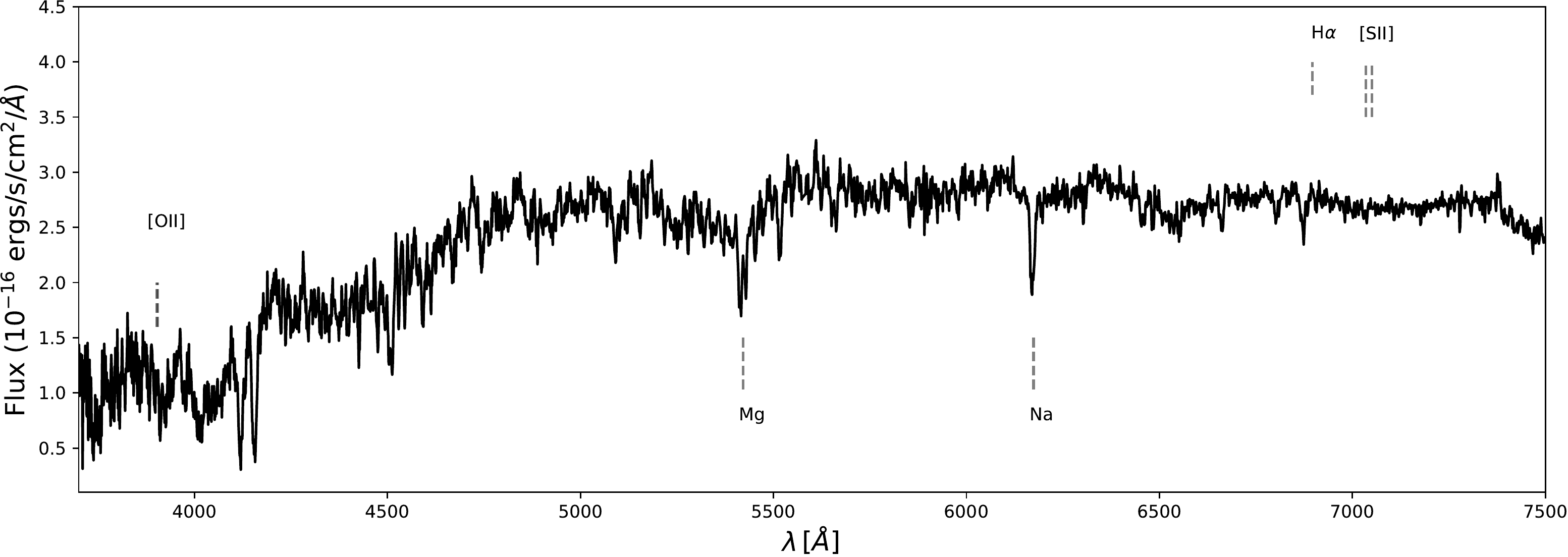}

\caption{Continued. Here we show the integrated spectrum for X3 and its corresponding host galaxy.}
\end{center}
\end{figure*}

\begin{figure*}
\begin{center}
\includegraphics[scale=0.5]{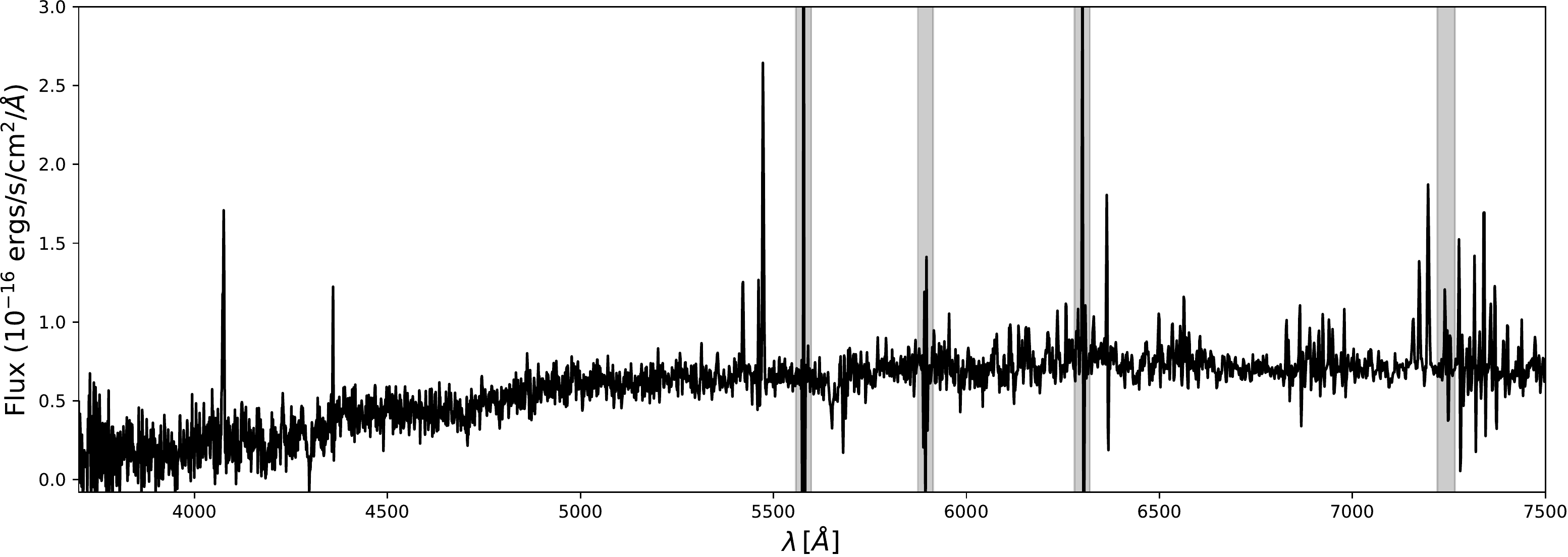}
\includegraphics[scale=0.5]{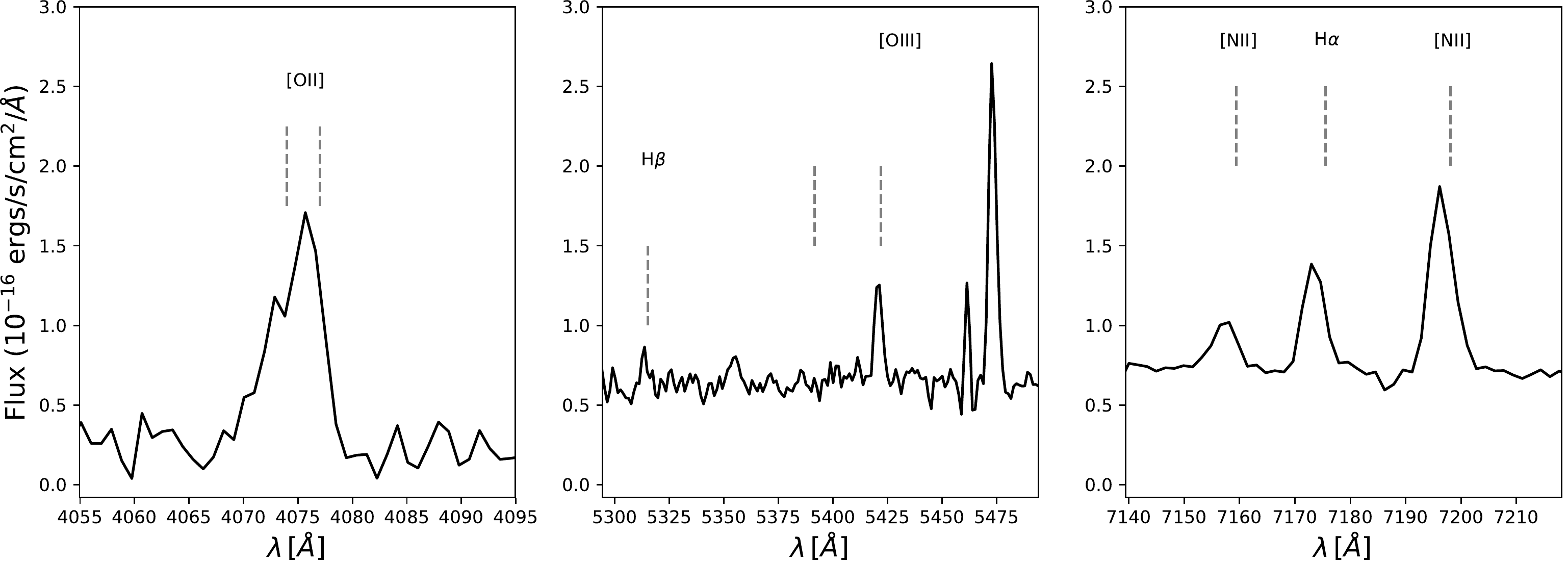}
\includegraphics[scale=0.5]{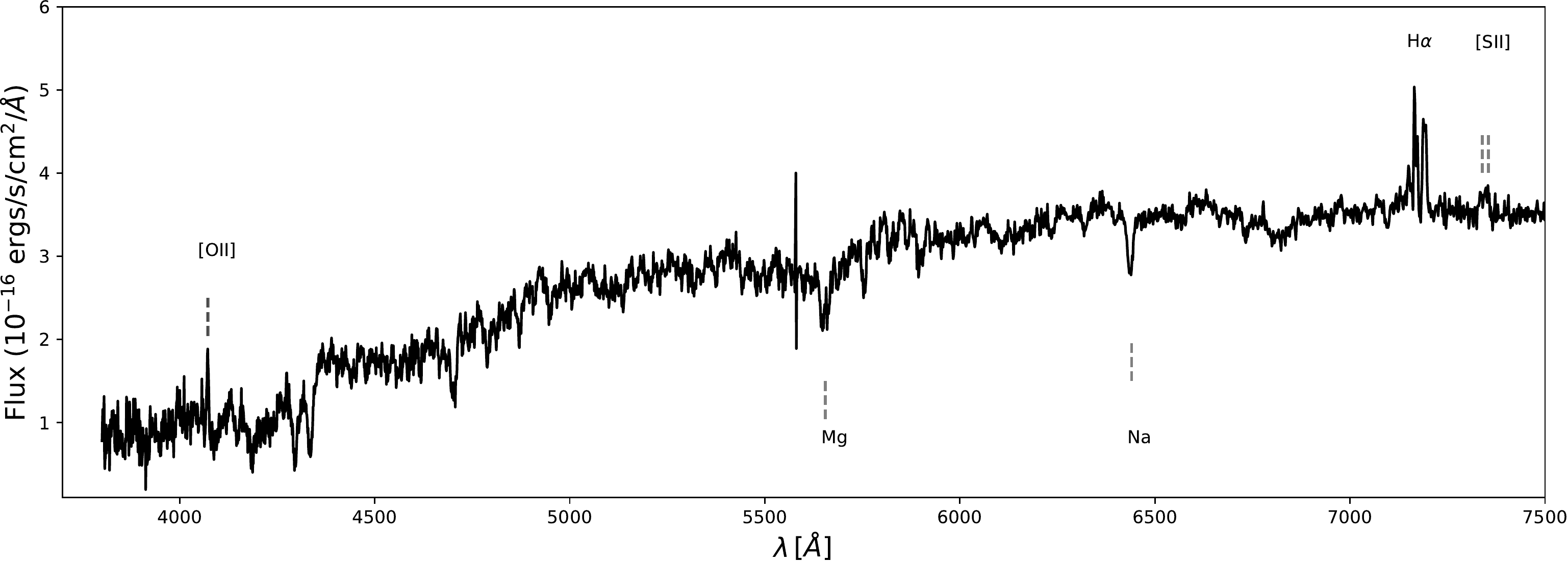}

\caption{Continued. Here we show the integrated spectrum for X4 and its corresponding host galaxy.}
\end{center}
\end{figure*}

\begin{figure*}
\begin{center}
\includegraphics[scale=0.5]{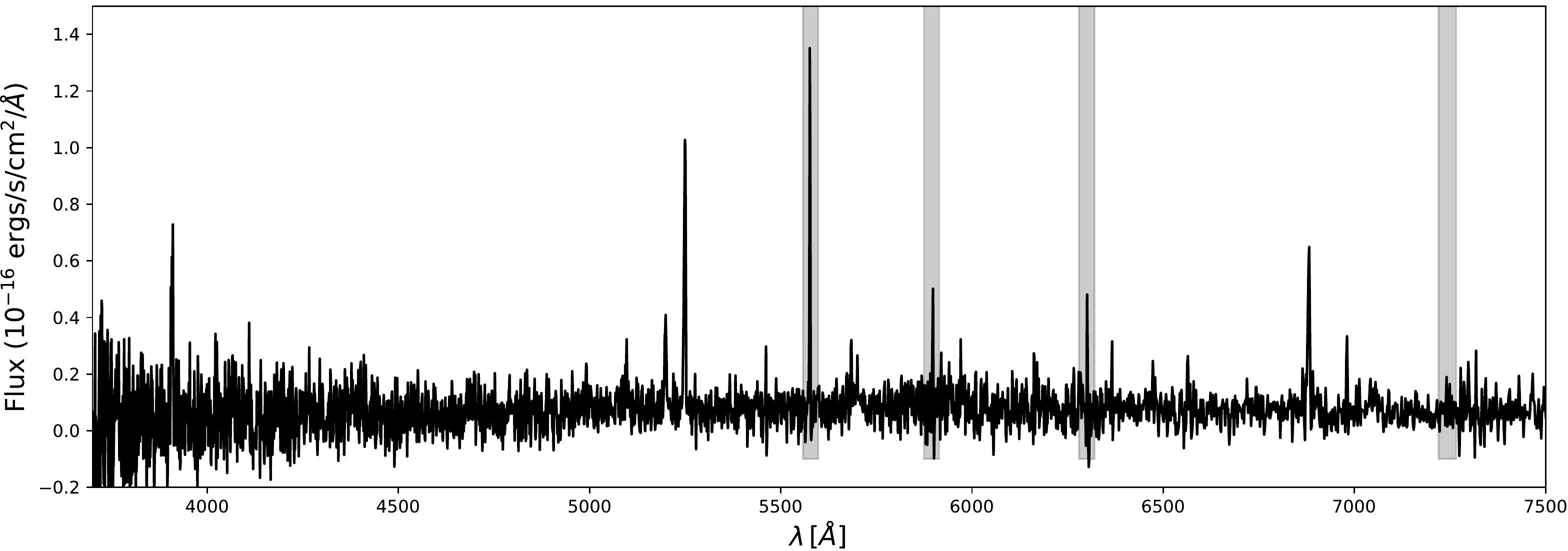}
\includegraphics[scale=0.5]{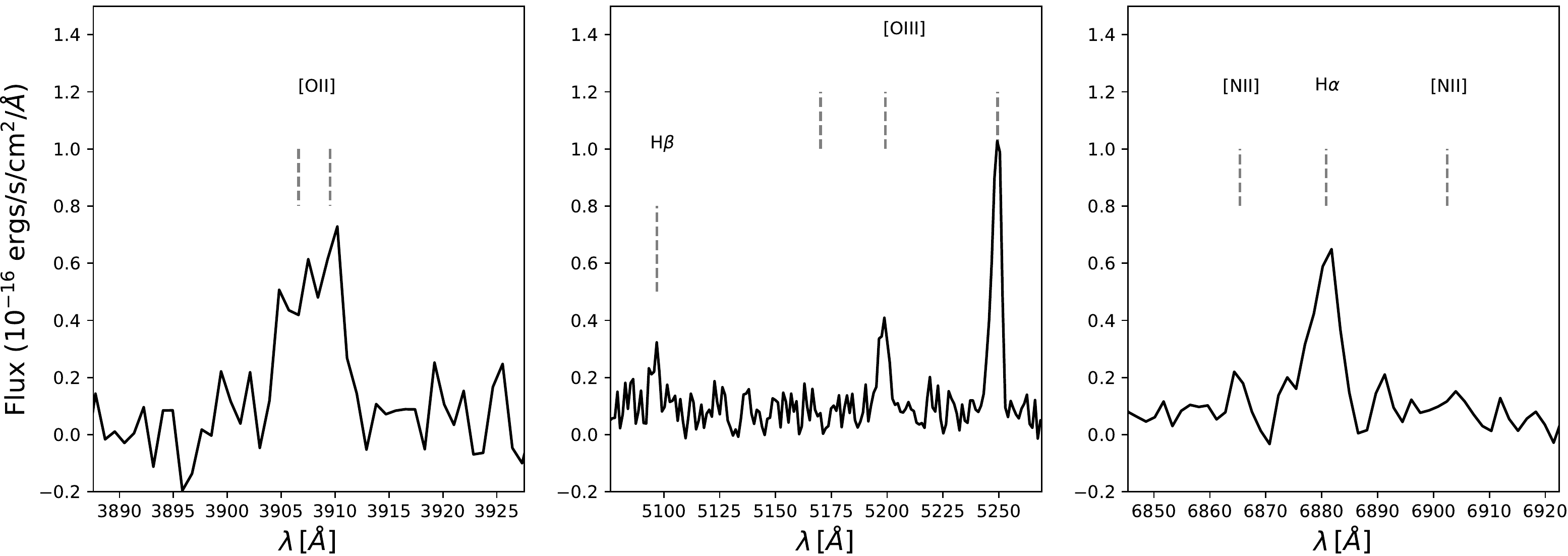}
\includegraphics[scale=0.5]{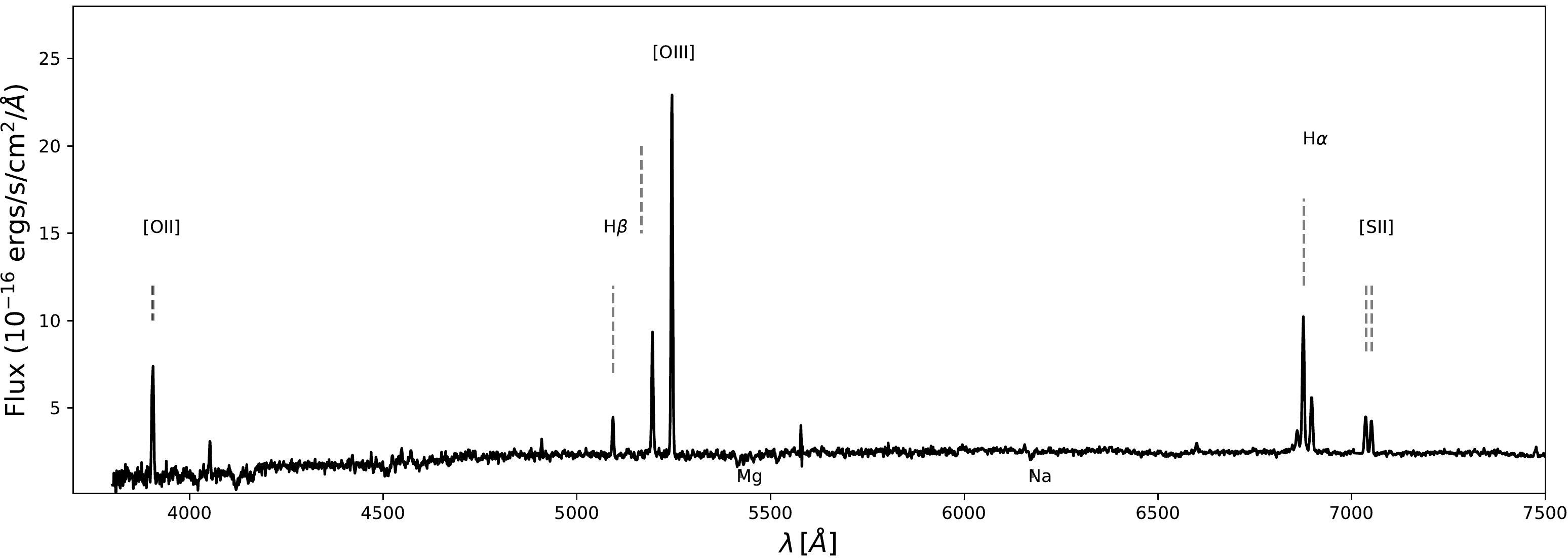} 

\caption{Continued. Here we show the integrated spectrum for X5 and its corresponding host galaxy.}\label{int_spec_5}
\end{center}
\end{figure*}

\section{List of artefacts found in the MaNGA datacube}\label{artefacts}
In our visual search for outlying \ha{} emitters in the entire data release 14 of the Sloan Digital Sky Survey (SDSS) IV MaNGA data release, we have found the following list of objects which are found to be artefacts. These artefacts are of two kinds: ones that appear like unresolved blobs (see Table \ref{blobs}) and ones that appear like extended sources. All the unresolved blobs appear like high signal-noise ratio emitters, with a different velocity component (see Fig. \ref{artefact} in the main text). However, a manual inspection of their spectra shows that their \ha{} FWHM is much narrower than the instrumental width. Additionally, they do not show line emission in other lines like H$\beta$, [OIII], [NII] which usually accompany strong H$\alpha$ emission. We believe that these artefacts are caused by some spurious low-intensity cosmic rays which are not correctly flagged by the MaNGA data reduction pipeline\footnote{\url{https://www.sdss.org/dr14/manga/manga-caveats/}}.

The extended emitters shows two types of artefacts, the first type is same as that of the unresolved blobs, which show a very narrow line width, the second kind show a strong \ha{} emission in the \piped{} output, however there is no emission line in the MaNGA datacube. Moreover, the integrated \ha{}/H$\beta$ emission line ratio is unrealistically higher than expected by any realistic assumptions about the nebular regions. We surmise that such artefacts could be due to incorrect fitting in the \piped{} pipeline. See Table \ref{extended} for the full list.

\begin{table*}
\begin{center}
\caption{List of artefacts in the MaNGA datacube which appear like unresolved blobs.}
\begin{tabular}{cccccccccc}
\hline \\
Obj ID &  RA & DEC & plate-ifu  & Host SDSS ObJ ID &RA (Host) &  DEC (Host) & z\\
        &     J2000 (hhmmss)    &  J2000 (ddmmss) &  & & J2000 (deg)  &                      J2000 (deg) & &  \\
\hline \\
B1 & 15:28:03.533 & +42:48:41.535 & 7443-9102 & 1237662501076336774 &15:28:03.32& +42:48:49.11 & 0.0918  \\
B2  & 03:12:46.534 & -01:01:16.964   & 8081-12701 & 1237666299480899906& 03:12:47.18& -01:01:22.27 & 0.0818\\
B3  & 03:44:31.591  & +00:00:14.240 & 8086-12704 & 1237663238741622872&03:44:31.59 & +00:00:23.94 & 0.1087 \\
B4  & 07:30:40.578 &+40:11:37.756  & 8131-6103 & 1237663530789503529 &07:30:40.96 & +40:11:30.39 & 0.1204 \\
B5  & 08:00:28.325 & +41:39:46.527 & 8143-6103 & 1237673705042935885 & 08:00:27.99 & +41:39:38.32 & 0.0438 \\
B6  & 07:51:14.337 & +28:09:05.193  & 8146-9102 & 1237657630579163404 &07:51:14.77 & +28:09:13.01 & 0.0526 \\
B7  & 07:58:25.691 &+27:09:44.698  & 8149-12702 & 1237657773924417752 & 07:58:26.21 & +27:09:41.50 & 0.0473\\
B8  & 07:49:38.378 & +48:30:53.394 & 8239-9102 & 1237663786879942876 & 07:49:37.25 & +48:30:56.46 & 0.0229 \\
B9  & 09:07:18.893 & +41:23:18.897  & 8247-6102 & 1237657775543746865 & 09:07:18.10 & +41:23:18.33 & 0.0273\\
B10  & 11:08:01.255 & +46:51:09.80  & 8257-12703 & 1237660636535652531 &11:08:00.70 & +46:51:03.66 & 0.0252 \\
B11  & 12:00:42.109 & +43:21:14.227  & 8259-12705 & 1237661968502358077 &12:00:41.45 & +43:21:16.47 & 0.1131 \\
B12  & 12:17:32.214 & +43:45:25.412  & 8262-3704 & 1237661871329640526 &12:17:32.74 & +43:45:26.50 & 0.0244 \\
B13  & 14:20:04.917 & +47:07:08.663  & 8326-6102 & 1237661957225119836 &14:20:04.29 & +47:07:16.82 & 0.07038\\
B14  & 13:58:33.421 & +41:43:32.770 & 8332-3704 & 1237661850403602575 &13:58:33.54 & +41:43:27.00 & 0.0431 \\
B15  & 11:31:20.193 & +21:25:41.810  & 8338-12705 & 1237667734503882923 &11:31:20.04 & +21:25:28.85 & 0.1337\\
B16  & 13:46:48.895 & +39:04:48.592 & 8447-12702 & 1237664296371421207 & 13:46:49.38 & +39:05:01.32 & 0.0618 \\
B17  & 09:49:31.512 & +42:07:34.148  & 8459-12701 & 1237660637066100922 &09:49:30.96 & +42:07:48.99 & 0.0721\\
B18  & 15:55:47.193 & +56:07:37.761 & 8481-3703 & 1237651539797737641 &15:55:46.82 & +56:07:31.92& 0.0423 \\
B19  & 16:00:44.508 & +53:46:24.932  & 8481-3704 & 1237651538724978813 &16:00:44.81 & +53:46:31.93 & 0.1104 \\
B20  & 16:27:45.899 & +44:10:29.753  & 8484-9101 &  1237655128765235356 &16:27:46.29 & +44:10:39.00 & 0.1398\\
B21  & 15:31:17.835 & +45:25:11.211  & 8551-12702 & 1237661386000826593 &15:31:17.56 & +45:24:59.53 & 0.07051\\
B22  & 07:39:07.686 & +41:23:33.537  & 8566-6102 & 1237673310428856407 &07:39:07.71 & +41:23:39.42 & 0.0986\\
B23  & 10:22:09.520 & +38:31:01.478 & 8568-12704 & 1237661137960632446 &10:22:10.31 & +38:31:04.17 & 0.0533 \\
B24  & 16:32:32.982 & +39:07:39.957  & 8603-12701 & 1237659326566564084 &16:32:33.73 & +39:07:51.74 & 0.1306 \\
B25  &21:10:15.857 & +10:24:20.440  & 8618-12705 & 1237653008120610948 &21:10:16.51 & +10:24:31.63 & 0.1186 \\
B26  & 08:10:42.094 & +48:55:06.267  & 8720-6104 & 1237651496296382666 &08:10:42.56 & +48:55:07.58 & 0.03864 \\
B27  & 08:52:41.457 & +54:25:57.881  & 8724-9101 & 1237651192432885982 &08:52:40.82 & +54:26:03.09 & 0.0386 \\
B28  & 10:25:42.860 & +36:01:31.483  & 8943-12704 & 1237662224594436134 &10:25:43.87 & +36:01:24.98 & 0.0546\\
B29  & 14:48:51.586 & +30:33:43.129  & 9002-12701 & 1237662696493547575 &14:48:50.84 & +30:33:51.74 & 0.061 \\
\hline
\hline \\

\end{tabular} \label{blobs}
\end{center}
Notes.\ Column (1) gives the object ID of the outlying \ha{} emitter,
column (2) and (3) gives the RA and DEC of the outlying \ha{} emitter respectively. Columns (4) and (5) give the MaNGA  plateifu name and SDSS DR14 object ID of the host galaxy respectively. And columns (6), (7) and (8) give the RA and DEC and SDSS spectroscopic redshift of the host galaxy respectively.  
\end{table*}

\begin{table*}
\begin{center}
\caption{List of artefacts in the MaNGA datacube with extended features.}
\begin{tabular}{cccccccccc}
\hline \\
Obj ID &  RA & DEC & plate-ifu  & Host SDSS ObJ ID &RA (Host) &  DEC (Host) & z\\
        &     J2000 (hhmmss)    &  J2000 (ddmmss) &  & & J2000 (deg)  &                      J2000 (deg) & &  \\
\hline \\
X6$^a$  & 08:20:52.800 & +18:01:04.896 & 8241-12702 & 1237667142860407201 &08:20:52.48 & +18:00:55.96 & 0.0441\\
X7$^b$  & 10:28:15.339 & +42:58:07.480 & 8253-12701 & 1237660634384761005 &10:28:14.10 & +42:58:05.32 & 0.0452 \\
X8$^a$ & 23:50:35.753 & -1:07:52.374  & 8655-12705 & 1237663275779227768 &23:50:36.42 & -01:07:40.97 & 0.0456 \\
X9$^a$ & 07:46:03.970 & +39:53:23.003  & 8713-12702 & 1237651192424890568 & 07:46:04.58 & +39:53:11.06  & 0.0418 \\
X10$^b$  & 12:59:12.968 & +27:46:32.067 & 8931-6103 & 1237667323797700611 & 12:59:13.49 & +27:46:28.53  & 0.0231 \\
X11$^a$  & 15:29:43.233 & +27:06:47.309  & 9042-12701 & 1237662697034613032 & 15:29:44.01 & +27:06:43.18  & 0.0457 \\
\hline
\hline \\

\end{tabular} \label{extended}
\end{center}
Notes.\ Column (1) gives the object ID of the outlying \ha{} emitter where X\#$^a$ represents artefact due to the \piped{} pipeline and X\#$^b$ represents artefact due to a spurious cosmic ray,
column (2) and (3) gives the RA and DEC of the outlying \ha{} emitter respectively. Column (4) and (5) gives the MaNGA  plateifu name and SDSS DR14 object ID of the host galaxy respectively. And column (6), (7) and (8) gives the RA and DEC and SDSS spectroscopic redshift of the host galaxy respectively.  
\end{table*}

\end{document}